\documentclass[12pt]{article}
\usepackage{epsfig,cite,amsmath,feynarts,axodraw}

\input paperdef

\evensidemargin \oddsidemargin
\makeatletter
\renewcommand\section{\@startsection {section}{1}{\z@}%
                           {-3.5ex \@plus -1ex \@minus -.2ex}%
                           {2.3ex \@plus.2ex}%
                           {\mathversion{bold}\normalfont\Large\bfseries}}
\renewcommand\subsection{\@startsection{subsection}{2}{\z@}%
                           {-3.25ex\@plus -1ex \@minus -.2ex}%
                           {1.5ex \@plus .2ex}%
                           {\mathversion{bold}\normalfont\large\bfseries}}
\renewcommand\subsubsection{\@startsection{subsubsection}{3}{\z@}%
                           {-3.25ex\@plus -1ex \@minus -.2ex}%
                           {1.5ex \@plus .2ex}%
                           {\mathversion{bold}\normalfont\normalsize\bfseries}}
\makeatother

\newcommand{\nn}{\non}%

\def\_{\rule{.3em}{.15ex}}

\def\slash#1{\setbox0=\hbox{$#1$}#1\hskip-\wd0\dimen0=5pt\advance
       \dimen0 by-\ht0\advance\dimen0 by\dp0\lower0.5\dimen0\hbox
         to\wd0{\hss\sl/\/\hss}}

\allowdisplaybreaks

\begin{document}
\thispagestyle{empty}

\def\thefootnote{\fnsymbol{footnote}}

\begin{flushright}
DCPT/06/44\\
IPPP/06/22\\
MPP--2006--14\\
hep-ph/0604147\\
\end{flushright}

\vspace{1cm}

\begin{center}

{\large\sc {\bf Precise Prediction for \boldmath{$\MW$} in the MSSM }}
 
\vspace{0.4cm}

\vspace{1cm}

{\sc 
S.~Heinemeyer$^{1}$%
\footnote{email: Sven.Heinemeyer@cern.ch}%
, W.~Hollik$^{2}$%
\footnote{email: hollik@mppmu.mpg.de}%
, D.~St\"ockinger$^{3}$%
\footnote{email: Dominik.Stockinger@durham.ac.uk}%
, A.M.~Weber$^{2}$%
\footnote{email: Arne.Weber@mppmu.mpg.de}%
, G.~Weiglein$^{3}$%
\footnote{email: Georg.Weiglein@durham.ac.uk}
}

\vspace*{1cm}

{\sl
$^1$Depto.\ de F\'isica Te\'orica, Universidad de Zaragoza, 50009 Zaragoza,
Spain 

\vspace*{0.4cm}

$^2$Max-Planck-Institut f\"ur Physik (Werner-Heisenberg-Institut),\\ 
F\"ohringer Ring 6, D--80805 Munich, Germany 

\vspace*{0.4cm}

%

$^3$IPPP, University of Durham, Durham DH1 3LE, U.K.

}

\end{center}

\vspace*{0.2cm}
\begin{abstract}
We present the currently most accurate evaluation of the $W$~boson
mass, $\MW$, in the Minimal Supersymmetric Standard Model (MSSM). 
The full complex phase dependence at the one-loop level, all available
MSSM two-loop corrections as well as the full Standard Model result have been
included. We analyse the impact of the different sectors of the MSSM
at the one-loop level with a particular emphasis on the effect of the
complex phases. We discuss the prediction for $\MW$ based on all known
higher-order contributions in representative MSSM scenarios.
Furthermore we obtain an estimate of
the remaining theoretical uncertainty from unknown higher-order 
corrections.
\end{abstract}

\def\thefootnote{\arabic{footnote}}
\setcounter{page}{0}
\setcounter{footnote}{0}

\newpage


\section{Introduction}

The relation between the massive gauge-boson masses, $\MW$ and $\MZ$, in
terms of the Fermi constant, $\gf$, and the fine structure constant,
$\al$, is of central importance for testing the electroweak theory.
It is usually employed for predicting $\MW$ in the model under
consideration. This prediction can then be compared with the
corresponding experimental value. The current experimental accuracy for
$\MW$, obtained at LEP and 
the Tevatron, is $\de \MW = 30 \mev$ (0.04\%)~\cite{LEPEWWG,LEPEWWG2}. 
This experimental resolution provides a high sensitivity to quantum
effects involving the whole structure of a given model. The $\MW$--$\MZ$
interdependence is therefore an important tool for discriminating
between the Standard Model (SM) and any extension or alternative of it,
and for deriving indirect constraints on unknown parameters such as the 
masses of the SM Higgs boson or supersymmetric particles, see 
\citere{PomssmRep} for a recent review. 
Within the SM the confrontation of the theory prediction and
experimental result for the $W$~boson mass supplemented by the other
precision observables yields an indirect constraint on the Higgs-boson
mass, $\MH$, of $\MH = 89^{+42}_{-30} \gev$ with $\MH < 175 \gev$
at the 95\%~C.L.~\cite{LEPEWWG2}.
Within the Minimal Supersymmetric Standard Model (MSSM)~\cite{susy}
the electroweak precision observables exhibit a
certain preference for a relatively low scale of supersymmetric
particles, see e.g.\ \citeres{ehow3,ehow4}. 

The experimental precision on $\MW$ will further improve within the next
years. The Tevatron data will reduce the experimental error to about
$\de\MW = 20 \mev$~\cite{mwTev},
while at the LHC an accuracy of about $\de\MW = 15 \mev$~\cite{mwlhc}
is expected. At the GigaZ
option of a linear $e^+e^-$ collider, a precision of
$\de \MW=7 \mev$ can be achieved~\cite{mwgigaz,blueband}.

A precise theoretical prediction for $\MW$ in terms of the model
parameters is of utmost importance for present and future electroweak
precision tests. Within the SM, the complete one-loop~\cite{MWSM1L} and
two-loop~\cite{MWSMferm2L,MWSMbos2L,drSMgfals,deltarSMgfals,MWSM} 
results are known as well as
leading higher-order 
contributions~\cite{drSMgfals2,MWSMQCD3LII,drSMgf3mh0,drSMgf3,drSMgf3MH}.

The theoretical evaluation of $\MW$ within the MSSM is not as advanced as 
in the SM. 
So far, the one-loop contributions have been
evaluated completely~\cite{dr1lA,dr1lB,MWMSSM1LA,MWMSSM1LB,pierce},
restricting  
however to the special case of vanishing complex phases (contributions
to the $\rho$ parameter with non-vanishing complex phases in the scalar
top and bottom mass matrices have been
considered in \citere{kangkim}). At the \twol\ 
level, the leading $\oaas$ corrections~\cite{dr2lA,dr2lB} and, most
recently, the leading electroweak corrections of \order{\alt^2}, 
\order{\alt \alb}, \order{\alb^2}to $\De\rho$ have been
obtained~\cite{drMSSMal2B,drMSSMal2A}. 
Going beyond the minimal SUSY model and allowing for non-minimal
flavor violation the leading one-loop contributions are
known~\cite{delrhoNMFV}. 

In order to fully exploit the experimental precision for testing 
supersymmetry (SUSY)
and deriving constraints on the supersymmetric parameters,%
\footnote{A precise prediction for $\MW$ in the MSSM is also needed as 
a part of the ``SPA Convention and Project'', see \citere{spa}.}
it is
desirable to have a prediction of $\MW$ in the MSSM at the same level
of accuracy as in the SM. As a step into this direction, we perform in
this paper a 
complete one-loop calculation of all contributions to $\MW$ in the MSSM
with complex parameters (cMSSM),
taking into account for the first time the full phase dependence
and imposing no restrictions on the various soft SUSY-breaking parameters. 
We combine this result with the full set of higher-order contributions
in the SM and with all available corrections in the MSSM. In this way we
obtain the currently most complete result for $\MW$ in the MSSM. A
public computer code based on our result for $\MW$ is in preparation.

We analyse the numerical results for $\MW$ for various
scenarios in the unconstrained MSSM and for SPS benchmark
scenarios~\cite{sps}. The dependence of the result on the complex phases 
of the soft SUSY-breaking parameters is investigated. We 
estimate the remaining theoretical uncertainties
from unknown higher-order corrections.

The outline of the paper is as follows. 
In \refse{sec:basics} the basic relations needed for the 
prediction of $\MW$ are given and our conventions and notations for the
different SUSY sectors are defined. In \refse{sec:1Lcalc} 
the complete one-loop result $\MW$ including the full phase 
dependence of the complex MSSM parameters is obtained. The
incorporation of all available higher-order corrections in the SM and
the MSSM is described.
The numerical analysis is presented in \refse{sec:numanal}, where we
also estimate the remaining theoretical uncertainties in the prediction
for $\MW$.
We conclude with \refse{sec:conclusions}.


\section{Prediction for $M_W$ --  basic entries}
\label{sec:basics}

Muons decay to almost 100\% into 
$e \nu_{\mu} \bar{\nu}_e$~\cite{pdg}. 
Historically, this decay process was first described within
Fermi's effective theory. 
The muon decay rate is related to Fermi's constant, $G_\mu$, by the defining
equation
\begin{equation}
\label{muondecayrate}
\Gamma_\mu =  \frac{G_\mu^2 m_\mu^5}{192\pi^3}\,
            F(\frac{m_e^2}{m_\mu^2})
            \left(1+\frac{3}{5} \frac{m_\mu^2}{M_W^2} \right) 
            (1+\Delta_{\rm QED}),
\end{equation}
with $F(x)=1-8x-12x^2\ln x +8x^3-x^4$. By convention, 
the QED corrections
in the effective theory, $\Delta_{\rm QED}$, are included in
\refeq{muondecayrate} as well as the (numerically insignificant) term
$3 m_\mu^2/(5 \MW^2)$ arising from the tree-level $W$ propagator.
The precise measurement of the muon lifetime 
and the equivalently precise calculation of 
$\Delta_{\rm QED}$~\cite{delqol,delqtl}
thus provide
the accurate value 
\begin{equation}
G_{\mu}=(1.16637\pm0.00001\times 10^{-5}) \gev^{-2}.
\end{equation}
In the SM and in the MSSM,  $G_\mu$ is
determined as a function of the basic model parameters. The
corresponding relation can be written as follows,
\begin{equation}
\frac{G_{\mu}}{\sqrt{2}}=\frac{e^2}
      {8 \left(1-\frac{\MW^2}{\MZ^2}\right) \MW^2}\, (1+\De r).
\label{GFdeltarrelation}
\end{equation}
The quantity $\De r$ summarizes the non-QED quantum corrections,
since QED
quantum effects are already included in the definition of
$\gf$ according to \refeq{muondecayrate},
which makes the evaluation of $\De r$ insensitive to infrared
divergences. 
$\De r$ depends on all the model parameters, which enter through 
the virtual states of all particles in loop diagrams, 
\begin{equation}
\De r = \De r (\MW,\MZ,\mt,\alpha,\alpha_s,\ldots,X)
\label{MWit}
\end{equation}
with
\begin{eqnarray}
                  && X = \MHSM ~~(\SM) , \nn\\ 
                  && X= \Mh, \MH, \MA,  \MHp, \tb, \msusy, A_f,
    m_{\tilde\chi^{0,\pm}}, \dots~~\textup{(MSSM)} , \nn
\end{eqnarray}
and is hence a model-specific quantity.
For each specification of the parameter set $X$,
the basic relation (\ref{GFdeltarrelation})
is fulfilled only by one value of $M_W$. This value can be considered
the model-specific prediction of the $W$ mass, $M_W(X)$, 
providing a sensitive precision observable, with different results in 
the SM and in the MSSM. 
Comparing the prediction of $\MW$ with the experimental
measurement leads to stringent tests of these models. 
Within the SM one can put bounds on the
Higgs boson mass, which has so far not been directly measured
experimentally. In the MSSM, $\De r$ and $\MW$ are sensitive to all the
free parameters of the model, such as SUSY particle 
masses, mixing angles and couplings. The SUSY entries 
are briefly described below, 
thereby introducing also conventions and notations
for the subsequent discussion.


\bigskip 
\noindent
\ul{Sfermions:}

\noindent
The mass matrix for the two sfermions of a given 
flavour, in the $\sfl, \sfr$ basis, 
is given by 
\begin{align}
{\matr{ M}}_{\tilde f} =
\begin{pmatrix}
        M_L^2 + \mf^2 & \mf \; \Xf^* \\
        \mf \; \Xf    & M_R^2 + \mf^2
\end{pmatrix} ,
\label{squarkmassmatrix}
\end{align}
with
\begin{eqnarray}
M_L^2 &=& M_{\tilde F}^2 + \MZ^2 \CZb (I_3^f - Q_f \sw^2) \non \\
M_R^2 &=& M_{\tilde F'}^2 + \MZ^2 \CZb Q_f \sw^2 \\ \non
\Xf &=& A_f - \mu^* \{\CTb, \tb\} ,
\label{squarksoftSUSYbreaking}
\end{eqnarray}
where $\{\CTb, \tb\}$ applies for up- and down-type sfermions,
respectively, and $\sw^2 \equiv \sin^2\theta_{\rm W} = 1 - \MW^2/\MZ^2$.
In the Higgs and scalar fermion sector of the cMSSM, $N_f + 1$ phases are
present, one for each $A_f$ and one
for $\mu$, i.e.\ $N_f + 1$ new parameters appear. As an abbreviation,
\begin{equation}
\phi_{\Af} \equiv \arg(\Af) \, ,
\end{equation}
will be used.
As an independent parameter one can trade $\phi_{\Af}$ for 
$\phi_{\Xf} \equiv \arg(\Xf)$.
The sfermion mass eigenstates are obtained by the transformation
\begin{equation}
\VL \sfe \\ \sfz \VR = {\matr{U}_{\tilde{f}}}_{\sf} \VL \sfl \\ \sfr \VR,
\end{equation}
with a unitary matrix $\matr{U}_{\tilde{f}}$.  
The mass eigenvalues are given by
\begin{equation}
m_{\tilde f_{1,2}}^2 = \mf^2
  + \edz \KKL M_L^2 + M_R^2
           \mp \sqrt{( M_L^2 - M_R^2)^2 + 4 \mf^2 |\Xf|^2}~\KKR ,
\label{sfermmasses}
\end{equation}
and are independent of the phase of $\Xf$.

\bigskip
\noindent
\ul{Higgs bosons:}

\noindent
Contrary to the SM, in the MSSM two Higgs doublets
are required.
At the tree level, the Higgs sector can be described with the help of two  
independent parameters, usually chosen as 
the ratio of the two
vacuum expectation values,  
$\tb \equiv v_2/v_1$, and $\MA$, the mass of the $\cp$-odd $A$ boson.
Diagonalisation of the bilinear part of the Higgs potential,
i.e.\ the Higgs mass matrices, is performed via rotations
by angles $\al$ for the $\cp$-even part and 
$\be$ for the $\cp$-odd and the charged part.
The angle $\al$ is thereby determined through
\begin{equation}
\tan 2\al = \tan 2\be \; \frac{\MA^2 + \MZ^2}{\MA^2 - \MZ^2} ;
\qquad  -\frac{\pi}{2} < \al < 0~.
\label{alphaborn}
\end{equation}
One obtains five physical states, the neutral $\cp$-even Higgs bosons
$h, H$, the $\cp$-odd Higgs boson $A$, and the charged Higgs bosons
$H^\pm$. Furthermore there are three unphysical Goldstone boson states,
$G^0, G^\pm$.
At lowest order, the Higgs boson masses can be expressed
in terms of $\MZ, \MW$, and $\MA$,
\begin{eqnarray}
\label{mlh}
 M_{h,H}^2 &=& \edz \KKL \MA^2 + \MZ^2  \mp 
          \sqrt{(\MA^2 + \MZ^2)^2 - 4 \MA^2\MZ^2 \CQZb} \KKR, \non \\
\label{mhp}
\MHp^2 &=& \MA^2 + \MW^2 .
\end{eqnarray}
Higher-order corrections, especially to $\Mh$, however, are 
in general quite large.
Therefore we use the results for the Higgs boson masses 
as obtained from the code \fhtt~\cite{feynhiggs,mhiggsAEC,feynhiggs2.2};
also the angle $\alpha$ is replaced by the
effective mixing angle $\alpha_{\rm eff}$ in the improved
Born approximation~\cite{hff,eehZhA}.

\bigskip
\noindent
\ul{Charginos and neutralinos:}

\noindent
The physical masses of the charginos are determined by the matrix
\begin{align}
  \matr{X} =
  \begin{pmatrix}
    M_2 & \sqrt{2} \sinb \MW \\
    \sqrt{2} \cosb \MW & \mu
  \end{pmatrix},
\end{align}
which contains the soft breaking term $M_2$ and the Higgsino mass term $\mu$,
both of which may have complex values in the cMSSM. Their complex
phases are denoted by 
\begin{equation}
\phi_{M_2} \equiv  {\rm arg}\KL M_2 \KR ~ \ \ \textup{and}\ \
\phi_{\mu} \equiv {\rm arg}\KL \mu \KR . 
\end{equation}
The physical masses are denoted as $m_{\tilde{\chi}^\pm_{1,2}}$ and
are obtained by applying the diagonalisation matrices 
   ${{\matr U}_{{\tilde{\chi}}^\pm}}$ and 
   ${{\matr V}_{{\tilde{\chi}}^\pm}}$, 
\begin{equation}
{\bf U^\ast_{{\tilde{\chi}}^\pm}}  \matr{X} {{\matr
    {V}}^\dag_{{\tilde{\chi}}^\pm}} = 
  \begin{pmatrix}
    m_{\tilde{\chi}^\pm_1} & 0 \\
    0 &  m_{\tilde{\chi}^\pm_2}
  \end{pmatrix} \, .
\end{equation}

The situation is similar for the neutralino masses, 
which can be calculated from the mass matrix
($\sw = \sin\theta_{\rm w}$, $\cw = \cos\theta_{\rm w}$)
\begin{align}
  \matr{Y} =
  \begin{pmatrix}
    M_1                    & 0                & -\MZ \, \sw \cosb
    & \MZ \, \sw \sinb \\ 
    0                      & M_2              & \quad \MZ \, \cw \cosb
    & \MZ \, \cw \sinb \\ 
    -\MZ \, \sw \cosb      & \MZ \, \cw \cosb & 0
    & -\mu             \\ 
    \quad \MZ \, \sw \sinb & \MZ \, \cw \sinb & -\mu                    & 0
  \end{pmatrix}.
\end{align}
This symmetric matrix contains the additional complex soft-breaking
parameter $M_1$, where the complex phase of $M_1$ is given by
\begin{equation}
\phi_{M_1} \equiv  {\rm arg}\KL M_1 \KR.
\end{equation}

The physical masses are denoted as $m_{\tilde{\chi}^0_{1,2,3,4}}$ and
are obtained in a diagonalisation procedure using the matrix
$\matr{N}_{\tilde{\chi}^0}$. 
\begin{align}
\matr{N}_{\tilde{\chi}^0}^\ast \matr{Y} \matr{N}_{\tilde{\chi}^0}^\dag=
  \begin{pmatrix}
   m_{\tilde{\chi}^0_{1}}                    & 0                & 0  & 0 \\ 
    0                      &   m_{\tilde{\chi}^0_{2}}           & 0  & 0 \\ 
    0      &  0 &  m_{\tilde{\chi}^0_{3}}    & 0             \\ 
  0 & 0 & 0                    & m_{\tilde{\chi}^0_{4}} 
  \end{pmatrix}.
\end{align}
At the two-loop level also the gluino enters the calculation of
$\MW$. In our calculation below we will incorporate the full phase
dependence of the complex parameters at the one-loop level, while we
neglect the explicit dependence on the complex phases beyond the
one-loop order. 
Accordingly, we take the soft SUSY-breaking parameter associated with
the gluino mass, $M_3 \equiv \mgl$, which enters only at two-loop
order, to be real.


\section{Calculation of $\Delta r$} 
\label{sec:1Lcalc}

Our aim is to obtain a maximally precise and general prediction for
$\MW$ in the MSSM. So far the one-loop result  
has been known only for
the case of real SUSY parameters \cite{MWMSSM1LA,MWMSSM1LB}. In this
section, we evaluate the complete one-loop result in
the cMSSM with general, complex parameters
and describe the incorporation of higher-order terms.


\subsection{Complete one-loop result in the complex MSSM}
\label{subsubsec:oneloop}

Evaluation of the full one-loop results requires renormalisation of
the tree-level Lagrangian. This introduces a set of one-loop counter
terms, which contribute to the
muon decay amplitude, in addition to the genuine one-loop graphs. 
At one-loop order, $\De r$ can be expressed 
in terms of the $W$~boson self-energy, vertex corrections (``vertex''),
box diagrams (``box''), and counterterms 
for charge, mass, and field renormalisation
as follows, 
\begin{eqnarray}
\De r^{(\al)} &=&\frac{\Si_T^{WW}(0)}{\MW^2}-\frac{\de
  \MW^2}{\MW^2}+ 2 \frac{\de
  e}{e}-\frac{\cw^2}{\sw^2}\left(\frac{\de
  \MZ^2}{\MZ^2}-\frac{\de \MW^2}{\MW^2}\right)\nn\\ 
&&+\frac{1}{2}\de Z^e_L+\frac{1}{2}\de
  Z^{\nu_e}_L+\frac{1}{2}\de Z^\mu_L+\frac{1}{2}\de
  Z^{\nu_\mu}_L +({\rm vertex})+({\rm box}) , 
\label{deltar1LCT}
\end{eqnarray}
where $\Si_T(q^2)$ denotes the transverse part of a vector boson
self-energy.
The leading contributions to $\De r^{(\al)}$ arise from the
renormalisation of the electric charge and the weak mixing angle (the
last two terms in the first line of \refeq{deltar1LCT}).
The former receives large fermionic contributions from the shift in the
fine structure constant due to light fermions, 
$\De\al \propto \log (m_f/\MZ)$,
a pure SM contribution. The latter involves the leading universal
corrections induced by the mass splitting between fields in an isospin
doublet~\cite{rho},
\begin{equation}
\De\rho = \frac{\Si_T^{ZZ}(0)}{\MZ^2} - \frac{\Si_T^{WW}(0)}{\MW^2} .
\label{eq:delrho}
\end{equation}
In the SM $\De\rho$ reduces to the well-known quadratic term in the
top-quark mass if the masses of the light fermions are neglected. 
In the MSSM $\De\rho$ receives additional sfermion contributions, in
particular from the squarks of the third generation.
The one-loop result for $\De r$ expressed in terms of $\De\al$ and
$\De\rho$ reads
\begin{equation}
\De r^{{(\al)}} = \De \al - \frac{\cw^2}{\sw^2}\De\rho 
                           + \De r_{\rm rem}^{(\al)},
\end{equation}
where $\De r_{\rm rem}^{(\al)}$ summarizes the remainder terms in 
\refeq{deltar1LCT}.

Throughout our calculation the on-shell
renormalisation scheme is applied. In this scheme renormalisation
conditions are imposed such that the particle masses are the poles of
the propagators and the fields  are renormalised by requiring unity
residues of the poles. These conditions ensure that the $\MW$--$\MZ$
interdependence given by \refeq{GFdeltarrelation} is a relation between
the physical masses of the two gauge bosons. Applying the
renormalisation conditions, the counterterms in
\refeq{deltar1LCT} can be expressed in terms of self-energies, and
eq.~(\ref{deltar1LCT}) turns into
\begin{eqnarray}
\De r ^{(\alpha)}
&=&\frac{\Si_T^{WW}(0)}{\MW^2}+({\rm vertex})+({\rm box})-
\frac{\re \Si_T^{WW}(\MW^2)}{\MW^2}
\nn\\&&
+\Big[\frac{\partial\Si^{\ga\ga}_T  
    (k^2)}{\partial
    k^2}\Big]_{k^2=0}-\frac{\sw}{\cw}\frac{\Si^{\ga Z}_T(0)}{\MZ^2}  
-\frac{\cw^2}{\sw^2}\,  \re \left[\frac{\Si^{ZZ}_T(\MZ^2)}{\MZ^2}
-\frac{\Si^{WW}_T(\MW^2)}{\MW^2}\right]\nn\\[0.2cm] 
&&- \Si^{e}_L(0) - \Si^{\mu}_L(0)
   - \Si^{\nu_{e}}_L(0) - \Si^{\nu_{\mu}}_L(0)\, ,    
\label{deltar1LOS}
\end{eqnarray}
where $\Si_L(q^2)$ denotes
the left-handed part of a fermion self-energy.
The electron and muon masses are neglected in the fermion
field renormalisation constants, which is possible since the 
only mass-singular virtual photon contribution is already contained
in $\Delta_{\rm QED}$ of eq.~(\ref{muondecayrate}) 
and is not part of $\De r$.

\begin{figure}[htb!]
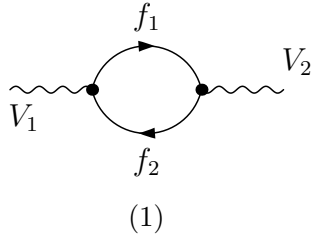

\centerline{
\unitlength=1.cm%
\begin{feynartspicture}(7,4)(1,1)
\FADiagram{$(1)$}
\FAProp(0.,10.)(6.,10.)(0.,){/Sine}{0}
\FALabel(1.,8.93)[t]{$V_1$}
\FAProp(20.,10.)(14.,10.)(0.,){/Sine}{0}
\FALabel(21.,11.07)[b]{$V_2$}
\FAProp(6.,10.)(14.,10.)(0.8,){/Straight}{-1}
\FALabel(10.,5.73)[t]{$f_2$}
\FAProp(6.,10.)(14.,10.)(-0.8,){/Straight}{1}
\FALabel(10.,14.27)[b]{$f_1$}
\FAVert(6.,10.){0}
\FAVert(14.,10.){0}
\end{feynartspicture}
}
\vspace{-1em}
\caption{Generic lepton and quark one-loop diagram
  contributing to $\De r$ via gauge-boson self-energies;
$V_1, V_2 = \ga, Z, W^\pm$, $f_1, f_2 = \nu, l, u, d$. 
} 
\label{fig:SMFermoneloop}
\end{figure}

\begin{figure}[htb!]
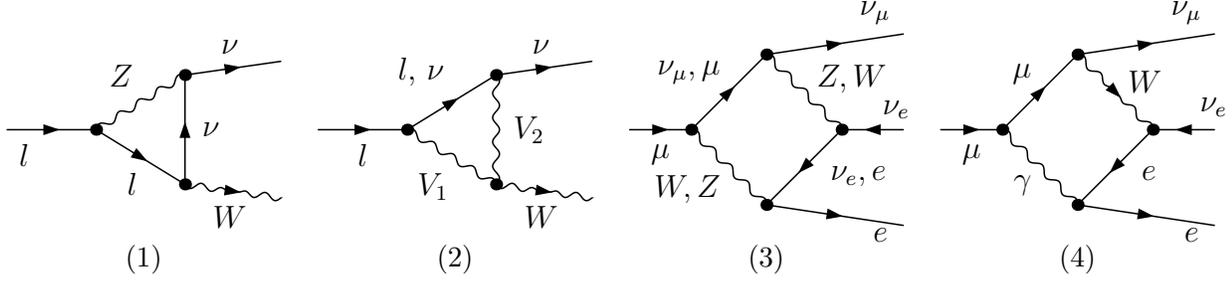

\centerline{
\unitlength=1.cm%
\begin{feynartspicture}(4,4)(1,1)
\FADiagram{(1)}
\FAProp(0.,10.)(6.5,10.)(0.,){/Straight}{1}
\FALabel(1.25,8.93)[t]{$l$}
\FAProp(20.,15.)(13.,14.)(0.,){/Straight}{-1}
\FALabel(16.2808,15.5544)[b]{$\nu$}
\FAProp(20.,5.)(13.,6.)(0.,){/Sine}{-1}
\FALabel(16.2808,4.44558)[t]{$W$}
\FAProp(6.5,10.)(13.,14.)(0.,){/Sine}{0}
\FALabel(9.20801,13.1807)[br]{$Z$}
\FAProp(6.5,10.)(13.,6.)(0.,){/Straight}{1}
\FALabel(9.5,6.81927)[tr]{$l$}
\FAProp(13.,14.)(13.,6.)(0.,){/Straight}{-1}
\FALabel(14.274,10.)[l]{$\nu$}
\FAVert(6.5,10.){0}
\FAVert(13.,14.){0}
\FAVert(13.,6.){0}
\end{feynartspicture}
%
%
%
%
\begin{feynartspicture}(4,4)(1,1)
\FADiagram{(2)}
\FAProp(0.,10.)(6.5,10.)(0.,){/Straight}{1}
\FALabel(3.25,8.93)[t]{$l$}
\FAProp(20.,15.)(13.,14.)(0.,){/Straight}{-1}
\FALabel(16.2808,15.5544)[b]{$\nu$}
\FAProp(20.,5.)(13.,6.)(0.,){/Sine}{-1}
\FALabel(16.2808,4.44558)[t]{$W$}
\FAProp(6.5,10.)(13.,14.)(0.,){/Straight}{1}
\FALabel(9.20801,13.1807)[br]{$l$, $\nu$}
\FAProp(6.5,10.)(13.,6.)(0.,){/Sine}{0}
\FALabel(9.5,6.81927)[tr]{$V_1$}
\FAProp(13.,14.)(13.,6.)(0.,){/Sine}{0}
\FALabel(14.274,10.)[l]{$V_2$}
\FAVert(6.5,10.){0}
\FAVert(13.,14.){0}
\FAVert(13.,6.){0}
\end{feynartspicture}
%
%
%
%
\begin{feynartspicture}(4,4)(1,1)
\FADiagram{(3)}
\FAProp(0.,10.)(4.5,10.)(0.,){/Straight}{1}
\FALabel(2.25,8.93)[t]{$\mu$}
\FAProp(20.,17.)(10.,15.5)(0.,){/Straight}{-1}
\FALabel(17.951,17.7651)[b]{$\nu_\mu$}
\FAProp(20.,3.)(10.,4.5)(0.,){/Straight}{-1}
\FALabel(18.3424,2.8)[t]{$e$}
\FAProp(20.,10.)(15.5,10.)(0.,){/Straight}{1}
\FALabel(19.4795,10.8041)[b]{$\nu_e$}
\FAProp(4.5,10.)(10.,15.5)(0.,){/Straight}{1}
\FALabel(6.63398,13.366)[br]{$\nu_\mu, \mu$}
\FAProp(4.5,10.)(10.,4.5)(0.,){/Sine}{0}
\FALabel(6.63398,6.63398)[tr]{$W,Z$}
\FAProp(10.,15.5)(15.5,10.)(0.,){/Sine}{0}
\FALabel(13.6968,12.6968)[bl]{$Z,W$}
\FAProp(10.,4.5)(15.5,10.)(0.,){/Straight}{-1}
\FALabel(14.6968,7.30322)[tl]{$\nu_e, e$}
\FAVert(4.5,10.){0}
\FAVert(10.,15.5){0}
\FAVert(10.,4.5){0}
\FAVert(15.5,10.){0}
\end{feynartspicture}
\begin{feynartspicture}(4,4)(1,1)
\FADiagram{(4)}
\FAProp(0.,10.)(4.5,10.)(0.,){/Straight}{1}
\FALabel(2.25,8.93)[t]{$\mu$}
\FAProp(20.,17.)(10.,15.5)(0.,){/Straight}{-1}
\FALabel(17.951,17.7651)[b]{$\nu_\mu$}
\FAProp(20.,3.)(10.,4.5)(0.,){/Straight}{-1}
\FALabel(18.3424,2.8)[t]{$e$}
\FAProp(20.,10.)(15.5,10.)(0.,){/Straight}{1}
\FALabel(19.951,11)[b]{$\nu_e$}
\FAProp(4.5,10.)(10.,15.5)(0.,){/Straight}{1}
\FALabel(6.63398,13.366)[br]{$\mu$}
\FAProp(4.5,10.)(10.,4.5)(0.,){/Sine}{0}
\FALabel(6.63398,6.63398)[tr]{$\gamma$}
\FAProp(10.,15.5)(15.5,10.)(0.,){/Sine}{1}
\FALabel(13.6968,12.6968)[bl]{$W$}
\FAProp(10.,4.5)(15.5,10.)(0.,){/Straight}{-1}
\FALabel(14.6968,7.30322)[tl]{$e$}
\FAVert(4.5,10.){0}
\FAVert(10.,15.5){0}
\FAVert(10.,4.5){0}
\FAVert(15.5,10.){0}
\end{feynartspicture}
}
\vspace{-1em}
\caption{Generic gauge-boson contributions to one-loop vertex and box 
 diagrams entering $\De r$ (the same diagrams as in the SM). The labels 
 $l$ and $\nu$ in the
 vertex diagrams stand for electron, muon and the corresponding
 neutrinos.}
\label{fig:GHMSSMboxvertexoneloop}
\end{figure}

\begin{figure}[htb!]
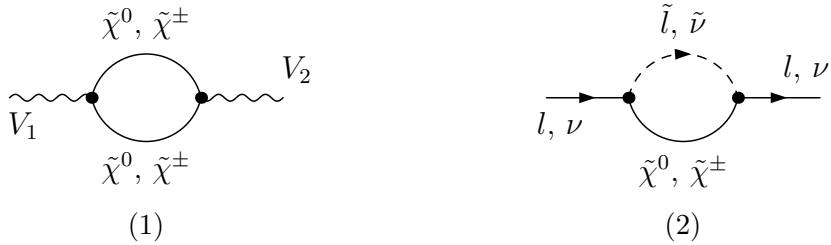

\centerline{
\unitlength=1.cm%
\begin{feynartspicture}(7,4)(1,1)
\FADiagram{(1)}
\FAProp(0.,10.)(6.,10.)(0.,){/Sine}{0}
\FALabel(1.,8.93)[t]{$V_1$}
\FAProp(20.,10.)(14.,10.)(0.,){/Sine}{0}
\FALabel(21.,11.07)[b]{$V_2$}
\FAProp(6.,10.)(14.,10.)(0.8,){/Straight}{0}
\FALabel(10.,5.98)[t]{$\tilde \chi^0$, $\tilde \chi^\pm$}
\FAProp(6.,10.)(14.,10.)(-0.8,){/Straight}{0}
\FALabel(10.,14.27)[b]{$\tilde \chi^0$, $\tilde \chi^\pm$}
\FAVert(6.,10.){0}
\FAVert(14.,10.){0}
\end{feynartspicture}
\begin{feynartspicture}(7,4)(1,1)
\FADiagram{(2)}
\FAProp(0.,10.)(6.,10.)(0.,){/Straight}{1}
\FALabel(1.,8.93)[t]{$l$, $\nu$}
\FAProp(20.,10.)(14.,10.)(0.,){/Straight}{-1}
\FALabel(19.,11.07)[b]{$l$, $\nu$}
\FAProp(6.,10.)(14.,10.)(0.8,){/Straight}{0}
\FALabel(10.,5.73)[t]{$\tilde \chi^0$, $\tilde \chi^\pm$}
\FAProp(6.,10.)(14.,10.)(-0.8,){/ScalarDash}{1}
\FALabel(10.,14.27)[b]{$\tilde{l}$, $\tilde{\nu}$}
\FAVert(6.,10.){0}
\FAVert(14.,10.){0}
\end{feynartspicture}
}
\vspace{-1em}
\caption{Generic neutralino/chargino contributions to gauge-boson
 (1) and fermion (2) self-energy diagrams. The labels 
 $l$, $\nu$ in the fermion self-energy diagram stand for electron, muon and
 their corresponding neutrinos, and the labels $\tilde l$, $\tilde \nu$ 
 indicate their respective superpartners.}
\label{fig:NCselfoneloop}
\end{figure}

\begin{figure}[htb!]
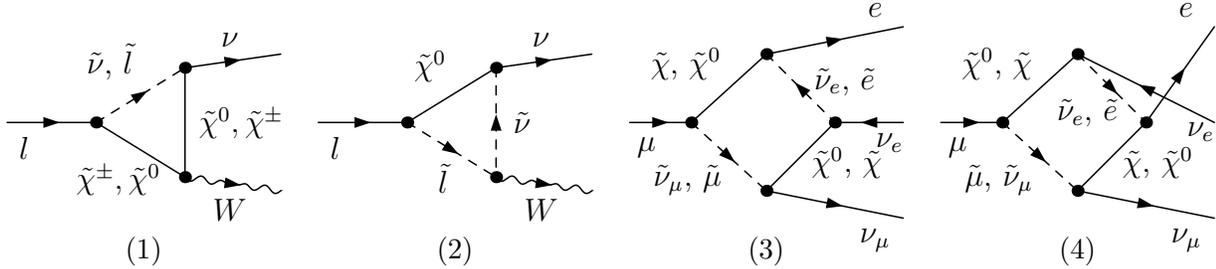

\centerline{
\unitlength=1.cm%
\begin{feynartspicture}(4.,4)(1,1)
\FADiagram{(1)}
\FAProp(0.,10.)(6.5,10.)(0.,){/Straight}{1}
\FALabel(1.25,8.93)[t]{$l$}
\FAProp(20.,15.)(13.,14.)(0.,){/Straight}{-1}
\FALabel(16.2808,15.5544)[b]{$\nu$}
\FAProp(20.,5.)(13.,6.)(0.,){/Sine}{-1}
\FALabel(16.2808,4.44558)[t]{$W$}
\FAProp(6.5,10.)(13.,14.)(0.,){/ScalarDash}{1}
\FALabel(9.20801,13.1807)[br]{$\tilde \nu$, $\tilde{l}$}
\FAProp(6.5,10.)(13.,6.)(0.,){/Straight}{0}
\FALabel(11.20801,7)[tr]{$\tilde \chi^\pm, \tilde \chi^0$}
\FAProp(13.,14.)(13.,6.)(0.,){/Straight}{0}
\FALabel(14.024,10.)[l]{$\tilde \chi^0, \tilde \chi^\pm$}
\FAVert(6.5,10.){0}
\FAVert(13.,14.){0}
\FAVert(13.,6.){0}
\end{feynartspicture}
\begin{feynartspicture}(4.,4)(1,1)
\FADiagram{(2)}
\FAProp(0.,10.)(6.5,10.)(0.,){/Straight}{1}
\FALabel(1.25,8.93)[t]{$l$}
\FAProp(20.,15.)(13.,14.)(0.,){/Straight}{-1}
\FALabel(16.2808,15.5544)[b]{$\nu$}
\FAProp(20.,5.)(13.,6.)(0.,){/Sine}{-1}
\FALabel(16.2808,4.44558)[t]{$W$}
\FAProp(6.5,10.)(13.,14.)(0.,){/Straight}{0}
\FALabel(9.33903,12.9678)[br]{$\tilde \chi^0$}
\FAProp(6.5,10.)(13.,6.)(0.,){/ScalarDash}{1}
\FALabel(9.5,7)[tr]{$\tilde{l}$}
\FAProp(13.,14.)(13.,6.)(0.,){/ScalarDash}{-1}
\FALabel(14.274,10.)[l]{$\tilde \nu$}
\FAVert(6.5,10.){0}
\FAVert(13.,14.){0}
\FAVert(13.,6.){0}
\end{feynartspicture}
\begin{feynartspicture}(4.,4)(1,1)
\FADiagram{(3)}
\FAProp(0.,10.)(4.5,10.)(0.,){/Straight}{1}
\FALabel(1.25,8.93)[t]{$\mu$}
\FAProp(20.,17.)(10.,15.)(0.,){/Straight}{-1}
\FALabel(17.951,17.7651)[b]{$e$}
\FAProp(20.,10.)(15.,10.)(0.,){/Straight}{1}
\FALabel(19.,9.23)[t]{$\nu_e$}
\FAProp(20.,3.)(10.,5.)(0.,){/Straight}{-1}
\FALabel(17.951,2.23485)[t]{$\nu_{\mu}$}
\FAProp(4.5,10.)(10.,15.)(0.,){/Straight}{0}
\FALabel(6.68736,13.1669)[br]{$\tilde \chi$, $\tilde \chi^0$}
\FAProp(4.5,10.)(10.,5.)(0.,){/ScalarDash}{1}
\FALabel(6.72099,6.87009)[tr]{$\tilde \nu_\mu$, $\tilde \mu$}
\FAProp(10.,15.)(15.,10.)(0.,){/ScalarDash}{-1}
\FALabel(13.52,12.02)[bl]{$\tilde \nu_e$, $\tilde e$}
\FAProp(15.,10.)(10.,5.)(0.,){/Straight}{0}
\FALabel(13.3,8.2)[tl]{$\tilde \chi^0$, $\tilde \chi$}
\FAVert(4.5,10.){0}
\FAVert(10.,15.){0}
\FAVert(15.,10.){0}
\FAVert(10.,5.){0}
\end{feynartspicture}
\begin{feynartspicture}(4.,4)(1,1)
\FADiagram{(4)}
\FAProp(0.,10.)(4.5,10.)(0.,){/Straight}{1}
\FALabel(1.25,8.93)[t]{$\mu$}
\FAProp(20.,10.)(10.,15.)(0.,){/Straight}{1}
\FALabel(17.951,17.7651)[b]{$e$}
\FAProp(20.,17.)(15.,10.)(0.,){/Straight}{-1}
\FALabel(19.,10)[t]{$\nu_e$}
\FAProp(20.,3.)(10.,5.)(0.,){/Straight}{-1}
\FALabel(17.951,2.23485)[t]{$\nu_{\mu}$}
\FAProp(4.5,10.)(10.,15.)(0.,){/Straight}{0}
\FALabel(6.68736,13.1669)[br]{$\tilde \chi^0$, $\tilde \chi$}
\FAProp(4.5,10.)(10.,5.)(0.,){/ScalarDash}{1}
\FALabel(6.72099,6.87009)[tr]{$\tilde \mu$, $\tilde \nu_\mu$}
\FAProp(10.,15.)(15.,10.)(0.,){/ScalarDash}{1}
\FALabel(8.5,9.7)[bl]{$\tilde \nu_e$, $\tilde e$}
\FAProp(15.,10.)(10.,5.)(0.,){/Straight}{0}
\FALabel(13.3,8.2)[tl]{$\tilde \chi$, $\tilde \chi^0$}
\FAVert(4.5,10.){0}
\FAVert(10.,15.){0}
\FAVert(15.,10.){0}
\FAVert(10.,5.){0}
\end{feynartspicture}
}
\vspace{-1em}
\caption{Generic neutralino and chargino contributions to one-loop vertex 
  and box diagrams entering $\De r$. The labels 
 $l$, $\nu$ stand for electron, muon and
 their corresponding neutrinos, and the labels $\tilde l$, $\tilde \nu$ 
 indicate their respective superpartners.}
\label{fig:NCvertexboxoneloop}
\end{figure}

At the one-loop level one can divide the diagrams contributing to
$\Delta r$ into four classes:\\[-1.5em]
\begin{itemize}
\item[(i)] SM-like contributions of
quark and lepton loops in the gauge-boson self-energies, schematically
depicted in \reffi{fig:SMFermoneloop};

\item[(ii)] SUSY contributions of squark and
slepton loops in the gauge-boson self-energies, depicted
in \reffi{fig:SquarkSleptoneloop}; 

\item[(iii)] contributions from the Higgs and gauge boson
sector, which contain besides 
self-energies (\reffis{fig:GHMSSMSMlikeselfoneloop} and 
\ref{fig:GHMSSMSUSYselfoneloop}) 
also vertex and box graphs (\reffi{fig:GHMSSMboxvertexoneloop}).

\item[(iv)] SUSY contributions involving 
neutralinos and charginos
in self-energies, vertex graphs and box diagrams,
see \reffis{fig:NCselfoneloop} and
\ref{fig:NCvertexboxoneloop}.

\end{itemize}

%
%

%
\begin{figure}[htb!]
\centerline{
\unitlength=1.cm%
\begin{feynartspicture}(7,4)(1,1)
\FADiagram{$(1)$}
\FAProp(0.,10.)(10.,10.)(0.,){/Sine}{0}
\FALabel(3.,8.93)[t]{$V_1$}
\FAProp(20.,10.)(10.,10.)(0.,){/Sine}{0}
\FALabel(18.,8.93)[t]{$V_2$}
\FAProp(10.,10.)(10.,10.)(10.,15.5){/ScalarDash}{-1}
\FALabel(10.,16.57)[b]{$\tilde{f}$}
\FAVert(10.,10.){0}
\end{feynartspicture}
\begin{feynartspicture}(7,4)(1,1)
\FADiagram{$(2)$}
\FAProp(0.,10.)(6.,10.)(0.,){/Sine}{0}
\FALabel(1.,8.93)[t]{$V_1$}
\FAProp(20.,10.)(14.,10.)(0.,){/Sine}{0}
\FALabel(21.,11.07)[b]{$V_2$}
\FAProp(6.,10.)(14.,10.)(0.8,){/ScalarDash}{-1}
\FALabel(10.,5.73)[t]{$\tilde f_2$}
\FAProp(6.,10.)(14.,10.)(-0.8,){/ScalarDash}{1}
\FALabel(10.,14.27)[b]{$\tilde f_1$}
\FAVert(6.,10.){0}
\FAVert(14.,10.){0}
\end{feynartspicture}
}
\vspace{-1em}
\caption{Generic squark and slepton one-loop diagrams which
  contribute to $\De r$ via gauge-boson self-energies; 
$\tilde f, \tilde f_1, \tilde f_2 = \tilde \nu, \tilde l, \tilde u, \tilde d$. 
        } 
\label{fig:SquarkSleptoneloop}
\end{figure}

\begin{figure}[htb!]
\BC
\unitlength=1.cm%
\begin{feynartspicture}(5.2,4)(1,1)
\FADiagram{(1)}
\FAProp(0.,10.)(10.,10.)(0.,){/Sine}{0}
\FALabel(2.,8.93)[t]{$V_1$}
\FAProp(20.,10.)(10.,10.)(0.,){/Sine}{0}
\FALabel(19.,8.93)[t]{$V_2$}
\FAProp(10.,10.)(10.,10.)(10.,15.5){/Sine}{0}
\FALabel(10.,16.32)[b]{$V_3$}
\FAVert(10.,10.){0}
\end{feynartspicture}
%
\begin{feynartspicture}(5.2,4)(1,1)
\FADiagram{(2)}
\FAProp(0.,10.)(6.,10.)(0.,){/Sine}{0}
\FALabel(1.,8.93)[t]{$V_1$}
\FAProp(20.,10.)(14.,10.)(0.,){/Sine}{0}
\FALabel(21.,11.07)[b]{$V_2$}
\FAProp(6.,10.)(14.,10.)(0.8,){/Sine}{0}
\FALabel(10.,5.73)[t]{$V_4$}
\FAProp(6.,10.)(14.,10.)(-0.8,){/Sine}{0}
\FALabel(10.,14.27)[b]{$V_3$}
\FAVert(6.,10.){0}
\FAVert(14.,10.){0}
\end{feynartspicture}
\begin{feynartspicture}(5.2,4)(1,1)
\FADiagram{(3)}
\FAProp(0.,10.)(6.,10.)(0.,){/Straight}{1}
\FALabel(3.,8.93)[t]{$l$, $\nu$}
\FAProp(20.,10.)(14.,10.)(0.,){/Straight}{-1}
\FALabel(17.,11.07)[b]{$l$, $\nu$}
\FAProp(6.,10.)(14.,10.)(0.8,){/Straight}{1}
\FALabel(10.,5.73)[t]{$l$, $\nu$}
\FAProp(6.,10.)(14.,10.)(-0.8,){/Sine}{0}
\FALabel(10.,14.27)[b]{$V$}
\FAVert(6.,10.){0}
\FAVert(14.,10.){0}
\end{feynartspicture}
\EC
\vspace{-1.5em}
\caption{Generic gauge-boson contributions to one-loop gauge-boson and
  fermion self-energies entering $\De r$
  (the same diagrams as in the SM). The labels $l$
  and $\nu$ in the fermion self-energy diagram stand for electron,
  muon and the corresponding neutrinos. 
  } 
\label{fig:GHMSSMSMlikeselfoneloop}
\end{figure}
%
%
\begin{figure}[htb!]
\centerline{
\unitlength=1.cm%
\begin{feynartspicture}(5.2,4)(1,1)
\FADiagram{(1)}
\FAProp(0.,10.)(10.,10.)(0.,){/Sine}{0}
\FALabel(3.,8.93)[t]{$V_1$}
\FAProp(20.,10.)(10.,10.)(0.,){/Sine}{0}
\FALabel(17.,8.93)[t]{$V_2$}
\FAProp(10.,10.)(10.,10.)(10.,15.5){/ScalarDash}{0}
\FALabel(10.,16.32)[b]{$s$}
\FAVert(10.,10.){0}
\end{feynartspicture}
\begin{feynartspicture}(5.2,4)(1,1)
\FADiagram{(2)}
\FAProp(0.,10.)(6.,10.)(0.,){/Sine}{0}
\FALabel(1.,8.93)[t]{$V_1$}
\FAProp(20.,10.)(14.,10.)(0.,){/Sine}{0}
\FALabel(21.,11.07)[b]{$V_2$}
\FAProp(6.,10.)(14.,10.)(0.8,){/ScalarDash}{0}
\FALabel(10.,5.98)[t]{$s_2$}
\FAProp(6.,10.)(14.,10.)(-0.8,){/ScalarDash}{0}
\FALabel(10.,14.27)[b]{$s_1$}
\FAVert(6.,10.){0}
\FAVert(14.,10.){0}
\end{feynartspicture}
\begin{feynartspicture}(5.2,4)(1,1)
\FADiagram{(3)}
\FAProp(0.,10.)(6.,10.)(0.,){/Sine}{0}
\FALabel(1.,8.93)[t]{$V_1$}
\FAProp(20.,10.)(14.,10.)(0.,){/Sine}{0}
\FALabel(21.,11.07)[b]{$V_2$}
\FAProp(6.,10.)(14.,10.)(0.8,){/Sine}{0}
\FALabel(10.,5.98)[t]{$V_3$}
\FAProp(6.,10.)(14.,10.)(-0.8,){/ScalarDash}{0}
\FALabel(10.,14.27)[b]{$s$}
\FAVert(6.,10.){0}
\FAVert(14.,10.){0}
\end{feynartspicture}
}
\vspace{-1em}
\caption{Generic contributions of MSSM Higgs bosons and Goldstone
bosons to one-loop gauge-boson self-energies;
$s, s_1, s_2 = h, H, A, H^\pm, G^0, G^\pm$.
  } 
\label{fig:GHMSSMSUSYselfoneloop}
\end{figure}
%
%
%
%
%
%

The calculation of  
the one-loop diagrams was performed in the dimensional
regularisation~\cite{dreg} as well as the dimensional reduction
scheme~\cite{dred}. The analytical 
result for $\De r$ 
turned out to be independent of the choice of scheme. 

Technically, the calculation was done with support of the
{\tt Mathematica} packages {\tt FeynArts} \cite{feynarts,famssm},
{\tt OneCalc}~\cite{twocalc}, and {\tt FormCalc}~\cite{formcalc}. 
A peculiarity among the box diagrams is the graph with a virtual
photon, see diagram~(4) in \reffi{fig:GHMSSMboxvertexoneloop}.
Since QED corrections are accounted for in the Fermi model
definition of $\gf$, according to eq.~(\ref{muondecayrate}),
the corresponding diagram with a point-like $W$-propagator
$1/\MW^2$ has to be subtracted. 
This is the same procedure as in the SM
and leaves an IR-finite expression for $\De r$
(see \citere{MWSM1L} for the original analysis, a detailed discussion of
this point can be found in \citeres{MWSMferm2L,MWSMbos2L}).
After this subtraction, all external momenta and all lepton masses can be
neglected in the vertex and box diagrams, so that this set of Feynman 
diagrams reduces to one-loop vacuum integrals. 

In order to obtain the relevant contributions to $\De r$, the Born-level
amplitude needs to be factored out. This is straightforward for the SM
vertex and box graphs and also for the SUSY vertex corrections. 
Concerning the box contributions, diagrams like graph (3) of
\reffi{fig:GHMSSMboxvertexoneloop} directly yield a Born-like
structure (after simplifying the Dirac chain using, for instance, the 
Chisholm identity, 
$\ga_\mu\ga_\nu\ga_\rho = - i \epsilon_{\mu\nu\rho\si} \ga^\si\ga_5
+ g_{\mu\nu}\ga_\rho - g_{\mu\rho}\ga_\nu + g_{\nu\rho}\ga_\mu$), i.e.\
\begin{equation}
{\cal M}_{\textup{Born-like box}} = {\cal M}_{\textup{Born}} \cdot
\Delta r_{\textup{box}} .
\end{equation}
Here 
\begin{equation}
{\cal M}_{\textup{Born}} =
(\bar{u}_{\nu_\mu} \gamma^\rho\, \omega_{-}
u_{\mu})(\bar{u}_{e} \gamma_\rho\, \omega_{-} v_{\nu_e})\cdot \frac{2
  \pi \al}{\sw^2 \MW^2}
\end{equation}
is the tree-level matrix element in the limit where the momentum
exchange is neglected. Box graphs involving supersymmetric particles, 
on the other hand, yield a different spinor structure. The SUSY 
box graphs shown in \reffi{fig:NCvertexboxoneloop},
diagrams (3) and (4), can schematically be written 
as  
\BEA
\label{structuresusybox1}
{\cal M}_{\textup{SUSY box1}} &=& (\bar{u}_e \gamma^\rho\, \omega_{-}
u_{\mu})(\bar{u}_{\nu_\mu} \gamma_\rho\, \omega_{-} v_{\nu_e})\cdot
{b}_1^{(\al)}, \\
{\cal M}_{\textup{SUSY box2}} &=& (\bar{u}_{\nu_e} \omega_{-}
u_{\mu})(\bar{u}_{\nu_\mu} \omega_{+} v_{e})\cdot {b}_2^{(\al)}, 
\label{structuresusybox2}
\EEA
where \refeq{structuresusybox1} corresponds to diagram (3) and 
\refeq{structuresusybox2} to diagram (4). The SUSY box contributions to
$\Delta r$ can be extracted by applying the Fierz
identities
\BEA
(\bar{u}_e \gamma^\rho\, \omega_{-}
u_{\mu})(\bar{u}_{\nu_\mu} \gamma_\rho\, \omega_{-} v_{\nu_e}) &=&
- (\bar{u}_{\nu_\mu} \gamma^\rho\, \omega_{-}
u_{\mu})(\bar{u}_{e} \gamma_\rho\, \omega_{-} v_{\nu_e})
\label{eq:fierz1} \\
(\bar{u}_{\nu_e} \omega_{-}
u_{\mu})(\bar{u}_{\nu_\mu} \omega_{+} v_{e}) &=&
\frac{1}{2} (\bar{u}_{\nu_e}\gamma^\rho\,\omega_+
v_e)(\bar{u}_{\nu_{\mu}}\gamma_\rho\, \omega_- u_\mu) ~, 
\label{eq:fierz2}
\EEA
where \refeq{eq:fierz2} can be further manipulated using charge
conjugation transformations. This yields
\BE
\De r_{\textup{SUSY box1}} = 
-\frac{\sw^2 \MW^2}{2 \pi \al} {b}_1^{(\al)}, \quad
\De r_{\textup{SUSY box2}} = 
\frac{\sw^2 \MW^2}{4 \pi \al} {b}_2^{(\al)} .
\EE


\subsection{Incorporation of higher-order contributions} 
\label{sec:higherorders}

In order to make a reliable prediction for $\MW$ in the MSSM, the
incorporation of contributions beyond one-loop order is indispensable.
We now combine the one-loop result described in the previous section
with all known SM and MSSM higher-order contributions. In this way we
obtain the currently most accurate prediction for $\MW$ in the MSSM.


\subsubsection{Combining SM and MSSM contributions}
\label{subsec:SMandMSSM}

As mentioned before, the theoretical evaluation of $\MW$ (or $\De r$) in
the SM is 
significantly more advanced than in the MSSM. In order to obtain a
most
accurate prediction for $\MW$ (via $\De r$) within the MSSM it is
therefore useful to  
take all known SM corrections into account. This can be done by writing
the MSSM prediction for $\De r$ as
\begin{equation}
\De r^{\rm MSSM} = \De r^{\rm SM} + \De r^{{\rm MSSM}-{\rm SM}}~,
\label{eq:obsSMSUSY}
\end{equation}
where $\De r^{\rm SM}$ is the prediction in the SM 
and $\De r^{{\rm MSSM}-{\rm SM}}$ is the difference
between the MSSM and the SM prediction.

In order to obtain $\De r^{\rm MSSM}$ according to
\refeq{eq:obsSMSUSY} we
evaluate $\De r^{{\rm MSSM}-{\rm SM}}$ at the level of 
precision of the known MSSM corrections, while for $\De r^{\rm SM}$
we use the currently most advanced result in the SM including all known
higher-order corrections. As a consequence, $\De r^{\rm SM}$ 
takes into account higher-order contributions which are only
known for SM particles in the loop but not for their superpartners
(e.g.\ two-loop electroweak corrections beyond the leading Yukawa
contributions and three-loop corrections of
\order{\al\als^2}). 

It is obvious that the incorporation of all known SM contributions
according to \refeq{eq:obsSMSUSY} is advantageous in the decoupling
limit, where all superpartners 
are heavy and the Higgs sector becomes SM-like. In this case the 
second term in \refeq{eq:obsSMSUSY} goes to zero, so that the MSSM
result approaches the SM result with $\MHSM = \Mh$,
where $\Mh$ denotes the mass of the lightest $\cp$-even Higgs boson in
the MSSM.
For lower values of the scale of supersymmetry the contribution from 
supersymmetric particles in the loop can be of comparable size as
the known SM corrections. In view of the experimental bounds on the
masses of the supersymmetric particles (and the fact that supersymmetry
has to be broken), however, a complete cancellation between
the SM and supersymmetric contributions is not expected. Therefore it
seems appropriate to apply \refeq{eq:obsSMSUSY} also in this case
(see also the discussion in \citere{drMSSMal2B}).


\subsubsection{SM contributions}
\label{subsec:SMcontrib}

As mentioned above, within the SM the complete two-loop result has been
obtained for 
$\MW$~\cite{MWSMferm2L,MWSMbos2L,drSMgfals,deltarSMgfals,MWSM}. Besides 
the one-loop part of $\De r$~\cite{MWSM1L} it consists of the fermionic 
electroweak
two-loop contributions~\cite{MWSMferm2L}, the purely bosonic two-loop
contributions~\cite{MWSMbos2L} and the QCD corrections of
\order{\al\als}~\cite{drSMgfals,deltarSMgfals}. Higher-order QCD
corrections are known at
\order{\al\als^2}~\cite{drSMgfals2,MWSMQCD3LII}. 
Leading electroweak contributions of order
\order{\gf^2 \als \mt^4} and \order{\gf^3 \mt^6} that enter via the
quantity $\De\rho$ have been calculated in \citere{drSMgf3}.
Furthermore, purely fermionic three- and four-loop
contributions were obtained in \citere{MWpurelyferm}, but turned
out to be numerically very small due to accidental cancellations. 
The class of four-loop contributions obtained in 
\citere{SteinSchroed} give rise to a numerically negligible effect.

All numerically relevant contributions were combined in
\citere{MWSM}, and a compact expression for the total SM
result for $\MW$ was presented. This compact
expression approximates the full SM-result for $\MW$ to better than 
$0.5 \mev$ for Higgs masses ranging from 
$10 \gev \leq M_H \leq 1  \tev$, with the other parameters 
($\mt$, $\De\al^5_{\rm had}(\MZ)$, $\als(\MZ)$, $\MZ$)  varied within
$2 \sigma$ around their central experimental values. 
The contributions entering the result given in \citere{MWSM} 
can be written as 
\begin{equation}
\De r^{\rm SM} = \De r^{(\al)} + \De r^{(\al \als)} + 
\De r^{(\al^2)}_{\textup{ferm}}+ 
\De r^{(\al^2)}_{\textup{bos}} +  \De r^{(\al \als^2)}+ 
\De r^{(\gf^2 \als \mt^4)} +
\De r^{(\gf^3 \mt^6)}, 
\label{eq:deltarSM}
\end{equation} 
where we have suppressed the index ``SM'' on the right-hand side.


\subsubsection{MSSM two-loop contributions}
\label{subsubsec:twoloop}

\begin{figure}[htb!]
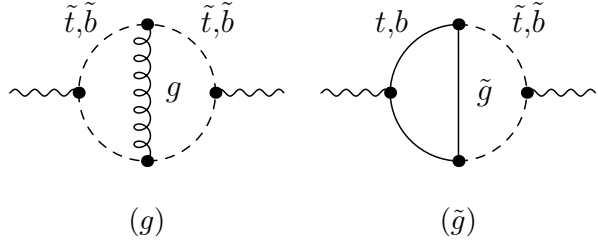

\centerline{
\unitlength=1.cm%
\begin{feynartspicture}(4,4)(1,1)
\FADiagram{$(g)$}
\FAVert(5,10){0}
\FAVert(15,10){0}
\FAProp(0,10)(5,10)(0.,){/Sine}{0}
\FAProp(15,10)(20,10)(0.,){/Sine}{0}
{
\FAProp(5,10)(15,10)(1.,){/ScalarDash}{0}
\FAProp(15,10)(5,10)(1.,){/ScalarDash}{0}
\FALabel(4,14)[b]{ ${\tilde{t}}$,${\tilde{b}}$}
\FALabel(14,14)[b]{ ${\tilde{t}}$,${\tilde{b}}$}}
\FAVert(10,15){0}
\FAVert(10,5){0}
\FAProp(10,5)(10,15)(0.,){/Cycles}{0}
\FALabel(11,10)[r]{$g$}
\end{feynartspicture}
\begin{feynartspicture}(4,4)(1,1)
\FADiagram{$(\tilde{g})$}
\FAVert(5,10){0}
\FAVert(15,10){0}
\FAProp(0,10)(5,10)(0.,){/Sine}{0}
\FAProp(15,10)(20,10)(0.,){/Sine}{0}
{
\FAProp(5,10)(10,15)(-.4,){/Straight}{0}
\FAProp(10,5)(5,10)(-.4,){/Straight}{0}
\FAProp(10,15)(15,10)(-.4,){/ScalarDash}{0}
\FAProp(10,5)(15,10)(.4,){/ScalarDash}{0}
\FALabel(4,14)[b]{ $t$,$b$}
\FALabel(14,14)[b]{ $\tilde{t}$,$\tilde{b}$}}
\FAVert(10,15){0}
\FAVert(10,5){0}
\FAProp(10,5)(10,15)(0.,){/Straight}{0}
\FALabel(11,10)[r]{$\tilde{g}$}
\end{feynartspicture}
}
\caption{Sample diagrams for the SUSY \order{\al \als} contributions to 
  $\De\rho$: $(g)$ squark loop with gluon
  exchange, $(\tilde{g})$ (s)quark loop with gluino exchange.}
\label{fig:samplediagramsQCD}
\end{figure}

The leading SUSY QCD corrections of \order{\al \als} entering via the
quantity $\De \rho$ arise from diagrams as
shown in \reffi{fig:samplediagramsQCD}, involving
gluon and gluino exchange in (s)top-(s)bottom loops.
These contributions were evaluated in \citere{dr2lA}. We have
incorporated this result into the term $\De r^{{\rm MSSM}-{\rm SM}}$
in \refeq{eq:obsSMSUSY}.

%

\begin{figure}[htb!]
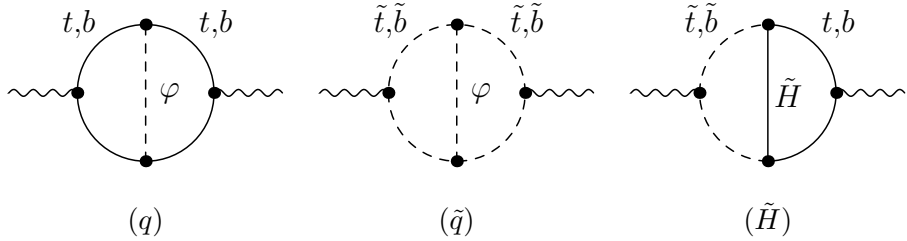

\centerline{
\unitlength=1.cm%
\begin{feynartspicture}(4,4)(1,1)
\FADiagram{$(q)$}
\FAVert(5,10){0}
\FAVert(15,10){0}
\FAProp(0,10)(5,10)(0.,){/Sine}{0}
\FAProp(15,10)(20,10)(0.,){/Sine}{0}
{
\FAProp(5,10)(15,10)(1.,){/Straight}{0}
\FAProp(15,10)(5,10)(1.,){/Straight}{0}
\FALabel(4,14)[b]{ ${t}$,${b}$}
\FALabel(14,14)[b]{ ${t}$,${b}$}}
\FAVert(10,15){0}
\FAVert(10,5){0}
\FAProp(10,5)(10,15)(0.,){/ScalarDash}{0}
\FALabel(11,10)[r]{$\varphi$}
\end{feynartspicture}
\unitlength=1.cm%
\begin{feynartspicture}(4,4)(1,1)
\FADiagram{$(\sq)$}
\FAVert(5,10){0}
\FAVert(15,10){0}
\FAProp(0,10)(5,10)(0.,){/Sine}{0}
\FAProp(15,10)(20,10)(0.,){/Sine}{0}
{
\FAProp(5,10)(15,10)(1.,){/ScalarDash}{0}
\FAProp(15,10)(5,10)(1.,){/ScalarDash}{0}
\FALabel(4,14)[b]{ $\tilde{t}$,$\tilde{b}$}
\FALabel(14,14)[b]{ $\tilde{t}$,$\tilde{b}$}}
\FAVert(10,15){0}
\FAVert(10,5){0}
\FAProp(10,5)(10,15)(0.,){/ScalarDash}{0}
\FALabel(11,10)[r]{$\varphi$}
\end{feynartspicture}
\unitlength=1.cm%
\begin{feynartspicture}(4,4)(1,1)
\FADiagram{$(\tilde{H})$}
\FAVert(5,10){0}
\FAVert(15,10){0}
\FAProp(0,10)(5,10)(0.,){/Sine}{0}
\FAProp(15,10)(20,10)(0.,){/Sine}{0}
{
\FAProp(5,10)(10,15)(-.4,){/ScalarDash}{0}
\FAProp(10,5)(5,10)(-.4,){/ScalarDash}{0}
\FAProp(10,15)(15,10)(-.4,){/Straight}{0}
\FAProp(10,5)(15,10)(.4,){/Straight}{0}
\FALabel(4,14)[b]{ $\tilde{t}$,$\tilde{b}$}
\FALabel(14,14)[b]{ ${t}$,${b}$}}
\FAVert(10,15){0}
\FAVert(10,5){0}
\FAProp(10,5)(10,15)(0.,){/Straight}{0}
\FALabel(11,10)[r]{$\tilde{H}$}
\end{feynartspicture}
}
\caption{Sample diagrams for the three classes of MSSM
  \order{\al_t^2}, \order{\al_b^2}, \order{\al_t \al_b} contributions
  to $\De\rho$: $(q)$ quark loop with Higgs
  exchange, $(\sq)$ squark loop with Higgs exchange, $(\tilde{H})$
  quark/squark loop with Higgsino exchange. $\varphi$ denotes Higgs and
  Goldstone boson exchange.}
\label{fig:samplediagramsYuk}
\end{figure}

Besides the \order{\al \als} contributions, recently also the
leading electroweak two-loop corrections of
\order{\al_t^2}, \order{\al_b^2} and \order{\al_t\al_b}
to $\De\rho$ have become available~\cite{drMSSMal2B}.
These two-loop Yukawa coupling contributions are due to
MSSM Higgs and Higgsino exchange in the (s)top-(s)bottom-loops, see
\reffi{fig:samplediagramsYuk}. 
In \citere{drMSSMal2B} the dependence of the
\order{\al_{t,b}^2} corrections on the lightest MSSM Higgs boson mass,
$\Mh$, has been analysed. Formally, at this order
the approximation $\Mh = 0$ would have to be employed.
However, it has been shown in \citere{drMSSMal2B} how a non-vanishing MSSM
Higgs boson mass can be consistently taken into account, including
higher-order corrections. Correspondingly we use the result of 
\citere{drMSSMal2B} for arbitrary $\Mh$ and employ the code
\fhtt~\cite{feynhiggs,mhiggsAEC,feynhiggs2.2} for the evaluation of the
MSSM Higgs sector parameters.

The irreducible supersymmetric two-loop contributions to $\De r$ 
discussed above need to be supplemented by the leading reducible
two-loop corrections. The latter can be obtained by expanding the 
resummation formula~\cite{CHJ}
\begin{equation}
1+\De r = \frac{1}{(1-\De \al)(1+\frac{\cw^2}{\sw^2}\De
  \rho)-\De r_{\textup{rem}}} 
\label{CHJresum}
\end{equation}
up to the two-loop order. At this order \refeq{CHJresum}
correctly incorporates terms of the type $(\De \al)^2$, $(\De
\rho)^2$, $\De \al \De\rho$ and $\De \al \De r_{\textup{rem}}$. In
this way we account for the leading terms of order \order{N_f^2
  \al^2}, where $N_f$ is the number of fermions. 

\bigskip
The final step is the inclusion of the complex MSSM parameters into 
the two-loop results. So far all two-loop results have been
obtained for real input parameters. 
We therefore approximate the two-loop result for a complex phase
$\phi$, $\MW^{\textup{full}}(\phi)$, by a simple interpolation
based on the full phase dependence at the one-loop level and
the known two-loop results for real
parameters, $\MW^{\textup{full}}(0)$, $\MW^{\textup{full}}(\pi)$, 
\BEA
\MW^{\rm full}(\phi) = \MW^{\rm 1L}(\phi) 
 &+& \left[ \MW^{\rm full}(0) - \MW^{\rm 1L}(0) \right] \times 
     \frac{1+\cos\phi}{2} \non \\
 &+& \left[ \MW^{\rm full}(\pi) - \MW^{\rm 1L}(\pi) \right] \times 
     \frac{1-\cos\phi}{2}~.
\label{phases2L}
\EEA
Here $\MW^{\rm 1L}(\phi)$ denotes the one-loop result, for which the
full phase dependence is known. The factors involving $\cos\phi$ 
ensure a smooth interpolation such that the known results 
$\MW^{\rm full}(0)$, $\MW^{\textup{full}}(\pi)$ are recovered for vanishing 
complex phase.
In \refse{sec:theounc} we estimate the 
uncertainty in the $\MW$ prediction due to this approximate inclusion
of the complex phase dependence at the two-loop level.


\subsection{Practical determination of $\De r$}
\label{deltar}

As can be seen from 
\refeqs{GFdeltarrelation},
(\ref{MWit}) $\MW$ is directly related to $\De r$,
which however depends on $\MW$ itself. In practical calculations we
therefore find $\MW$, using \refeq{GFdeltarrelation}, in an iterative 
procedure. According to our strategy outlined above we obtain $\MW$ 
from $\De r^{\rm SM}$ and $\De r^{{\rm MSSM}-{\rm SM}}$. In order to
make contact with the known SM result we use the  compact expression for
the total SM result for $\MW$ as given in \citere{MWSM}. This requires
some care in the iterative evaluation of $\MW$. Since the compact
expression for the SM result gives $\MW^{\textup{SM}}$,
its inversion yields $\De r^{\rm SM}(\MW^{\textup{SM}})$
rather than $\De r^{\rm SM}$ as function of the MSSM value of $\MW$,
which should be inserted in \refeqs{GFdeltarrelation}, (\ref{eq:obsSMSUSY}).

The desired expression for $\De r^{\rm SM}(\MW)$ is approximately given
by
\begin{eqnarray}
\De r^{\rm SM}(\MW) &\approx&
 \Big[\De r^{(\al)} + \De r^{(\al \als)} +
  \De r^{(\al \als^2)}
  +  \De r^{(\gf^2 \mt^4)} +  \De r^{(\al^2)}_{\textup{ferm, red}} \nonumber\\
&&{} + \De r^{(\gf^2 \als \mt^4)} +  \De r^{(\gf^3
    \mt^6)}\Big] (\MW)\nonumber\\
&&{} +
\Big[  \De   r^{(\al^2)}_{\textup{ferm, sublead}}
+ \De r^{(\al^2)}_{\textup{bos}}  \Big] (\MW^{\textup{SM}}),
\label{eq:delrsmmw}
\end{eqnarray}
where $\De r^{(\al^2)}_{\textup{ferm, red}}$ 
are reducible two-loop contributions arising from \refeq{CHJresum}.
All terms of the first two lines of the right-hand side of
\refeq{eq:delrsmmw} are known analytically. 
The numerically small contribution in the third line can therefore be
obtained as 
\begin{eqnarray}
\Big[  \De   r^{(\al^2)}_{\textup{ferm, sublead}}
+ \De r^{(\al^2)}_{\textup{bos}}  \Big] (\MW^{\textup{SM}}) &=&
\De r^{\rm SM}(\MW^{\textup{SM}})
-\Big[\De r^{(\al)} + \De r^{(\al \als)} +
  \De r^{(\al \als^2)} \nonumber\\
&& {} +  \De r^{(\gf^2 \mt^4)} +  \De r^{(\al^2)}_{\textup{ferm, red}}
      \nonumber\\
&& {} +  \De r^{(\gf^2 \als \mt^4)} +  \De r^{(\gf^3
    \mt^6)}\Big] (\MW^{\textup{SM}})
\end{eqnarray}
from $\De r^{\rm SM}(\MW^{\textup{SM}})$~\cite{MWSM}. Here
$\De   r^{(\al^2)}_{\textup{ferm, sublead}}$ denotes the subleading
fermionic two-loop terms beyond the leading $\mt^4$ term and the
reducible terms. 
In this way the correct $\MW$~dependence in $\De r$ is only neglected
in the subleading electroweak two-loop contributions 
$\De   r^{(\al^2)}_{\textup{ferm, sublead}}
+ \De r^{(\al^2)}_{\textup{bos}}$,
which are numerically small.
For $\De r^{(\al)}$ we use our recalculation, while
for  $\De r^{(\al \als)}$, ~$\De r^{(\al \als^2)}$, ~$\De r^{(\gf^2 \mt^4)}$,
~$\De r^{(\gf^2 \als \mt^4)}$ and ~$\De r^{(\gf^3 \mt^6)}$ 
we use compact expressions from\\
\citeres{MWSMQCD3LII,drSMgf3,DeltaRhoGF2MT4,drMSSMal2B}.  
Our final result for $\De r^{\rm MSSM}(\MW)$ reads
\begin{equation}
\De r^{\rm MSSM}(\MW) = \De r^{\rm SM}(\MW) + 
\De r^{{\rm MSSM}-{\rm SM}}(\MW)~,
\label{eq:DelrMSSMfinal}
\end{equation}
where $\De r^{\rm SM}(\MW)$ is given in \refeq{eq:delrsmmw} and
$\De r^{{\rm MSSM}-{\rm SM}}(\MW)$ is the difference between the full 
MSSM one-loop result as described in \refse{subsubsec:oneloop} and the
SM one-loop result supplemented by the higher-order supersymmetric
contributions specified in \refse{subsubsec:twoloop}. Inserting this 
expression into \refeq{GFdeltarrelation}
one can now calculate $\MW$ using
a standard iteration which is rapidly convergent.


\section{Numerical analysis}
\label{sec:numanal}

In the following subsections we present our numerical results.
First the impact of the one-loop contributions from the different
sectors of the MSSM is systematically analysed, and the dependence of the
result on the different masses and complex phases is studied in detail. 
As a second step we take into account all higher-order corrections and
discuss the full prediction for
the $W$~boson mass in the MSSM for a choice of sample scenarios.

For the numerical analysis the analytical results for $\De r$ and
$\MW$, which were calculated as described above,
were implemented into a Fortran program. Though built up from scratch,
for the calculation of the MSSM particle spectrum our code partially
relies on routines which are part of the {\tt FormCalc}
package~\cite{formcalc}. The Higgs sector parameters are obtained from
the program \fhtt~\cite{feynhiggs,mhiggsAEC,feynhiggs2.2}. 
This Fortran program for the calculation of precision observables within
the MSSM will be made publicly avaliable \cite{WaZOb}. 

If not stated otherwise, in the numerical analysis below for
simplicity we choose all soft 
SUSY-breaking parameters in the diagonal entries of the sfermion mass
matrices to be equal,
\begin{equation}
\msusy  \equiv M_{\tilde F} = M_{\tilde F'} = \ldots~,
\label{eq:Msusy}
\end{equation}
see \refeq{squarkmassmatrix}.
In the neutralino sector the GUT relation 
\begin{equation}
M_1 = \frac{5}{3} \frac{\sw^2}{\cw^2} M_2 
\label{eq:GUT}
\end{equation}
(for real values)
is often used to reduce
the number of free MSSM parameters.
We have kept $M_1$ as a free
parameter in our analytical calculations, but will use the GUT relation 
to specify $M_1$ for our numerical analysis if not stated otherwise.

We have fixed the SM input parameters as~%
\footnote{
For the results shown in \refse{subsec:oneloopres} we use
$\mt = 172.7 \pm 2.9 \gev$~\cite{mtexpnew}. For the comparison of
theoretical predictions with 
experimental data in \refse{subsec:MWpred} we use the most
up to date value $\mt = 172.5 \pm 2.3 \gev$, which has recently become
available \cite{newestmt}. 
}%
\BE 
\begin{aligned}
G_{\mu} &= 1.16637\times 10^{-5}, &
\MZ &= 91.1875 \gev, & 
\als(\MZ) &= 0.117 , \\
\alpha &= 1/137.03599911, &
\De \al^{(5)}_{\textup{had}} &= 0.02761 \textup{\cite{DeltaAlfaMartin}}, & 
\De \al_{\textup{lep}} &= 0.031498 \textup{\cite{DeltaAlfaStein}}, \\
\mt &= 172.5 \gev  \, &
\mb &= 4.7 \gev, &
m_\tau &= m_c = \ldots = 0 
\end{aligned} 
\label{eq:inputpars}
\EE
The complex phases appearing in the cMSSM are experimentally 
constrained by their
contribution to electric dipole moments of
heavy quarks~\cite{EDMDoink}, of the electron and 
the neutron (see \citeres{EDMrev2,EDMPilaftsis} and references therein), 
and of deuterium~\cite{EDMRitz}. While SM contributions enter 
only at the three-loop level, due to its
complex phases the cMSSM can contribute already at one-loop order.
Large phases in the first two generations of (s)fermions
can only be accomodated if these generations are assumed to be very
heavy~\cite{EDMheavy} or large cancellations occur~\cite{EDMmiracle},
see however the discussion in \citere{EDMrev1}. 
Accordingly (using the convention that $\phi_{M_2} =0$), in particular
the phase $\phi_\mu$ is tightly constrained~\cite{plehnix}, 
while the bounds on the phases of the third generation
trilinear couplings are much weaker.


\subsection{Analysis of parameter and phase dependence}
\label{subsec:oneloopres}

We begin by studying the impact of the one-loop contributions to 
$\De r$ from the various MSSM sectors, i.e.\ the sfermion sector, the
chargino and neutralino sector, as well as the gauge boson and Higgs
sector. In order to be able to analyse the
different sectors separately, we do not solve \refeq{GFdeltarrelation}
using our complete result for $\De r$, but we rather investigate 
the mass shift $\de\MW$ arising from changing $\De r$ by the amount 
$\De r^{\rm SUSY}$, 
\begin{equation}
\de \MW = -\frac{\MW^{\rm ref}}{2}  \frac{\sw^2}{\cw^2-\sw^2} 
\De r^{\rm SUSY} .
\label{deltaMW}
\end{equation}
Here $\De r^{\rm SUSY}$ represents the one-loop contribution from the
supersymmetric particles of the considered sector of the MSSM. We fix
the (in principle arbitrary) reference value for $\MW$ in \refeq{deltaMW}
to be $\MW^{\rm ref} = 80.425 \gev$. Our full result for $\MW$, which is 
determined from \refeq{GFdeltarrelation} in an iterative procedure, is
of course independent of this reference value.


\subsubsection{Sfermion sector dependence}
\label{subsec:sfsecdep}

We first investigate the influence of sfermion one-loop contributions, which
enter via the gauge-boson self-energy diagrams depicted in 
\reffi{fig:SquarkSleptoneloop}. The selectron and smuon contributions 
to the electron and muon field renormalisations shown in
\reffi{fig:NCselfoneloop} and to the vertex and box diagrams shown in
\reffi{fig:NCvertexboxoneloop} will be discussed as part of the
chargino and neutralino contributions in \refse{subsec:cndep}.

The leading one-loop SUSY contributions to $\De r$ arise
from the $\Stop/\Sbot$ doublet. Since the mass of the partner fermion
appears in the sfermion mass matrices, see \refeq{squarkmassmatrix},
a significant splitting between the diagonal entries can be induced in
the stop sector. The off-diagonal elements in the stop sector and 
for large $\tb$ also in the sbottom sector can furthermore give rise to
a large mixing between the two states of one flavour.

The complex parameters in the $\Stop/\Sbot$ sector are
$\mu$, $\At$ and $\Ab$. 
Neither the $\mu$ nor the $A$ parameters appear explicitly in the
couplings of the diagrams of \reffi{fig:SquarkSleptoneloop}.
They only enter via the absolute values and phases of
$X_{t,b}$, the off-diagonal entries of the squark mixing
matrices. We have checked at the analytical level that the phases
$\phi_{X_{t,b}}$ drop out entirely in the full one-loop calculation of
$\De r$ and have no influence on $\MW$. Hence, the phases and absolute
values of $\mu$, $\At$ and $\Ab$ enter the sfermion-loop
contributions (at one-loop order) only via
\begin{eqnarray}
|X_{t}|^2 &=&
|A_{t}|^2 +|\mu\CTb|^2-2 |A_{t}|\cdot|\mu|\CTb
\cos(\phi_{A_{t}}+\phi_{\mu})
\label{XtRel}
,\\
|X_{b}|^2 &=&
|A_{b}|^2 +|\mu\tb|^2-2 |A_{b}|\cdot|\mu| \tb
\cos(\phi_{A_{b}}+\phi_{\mu})
.
\label{XbRel}
\end{eqnarray}
In particular, the phases $\phi_{A_{t,b}}$ and $\phi_\mu$ only enter
in the combinations $(\phi_{A_{t,b}}+\phi_\mu)$ and only via modifications of
the squark masses and mixing angles.

\begin{figure}[htb!]
\begin{center}
\includegraphics[width=10cm,height=8.4cm]{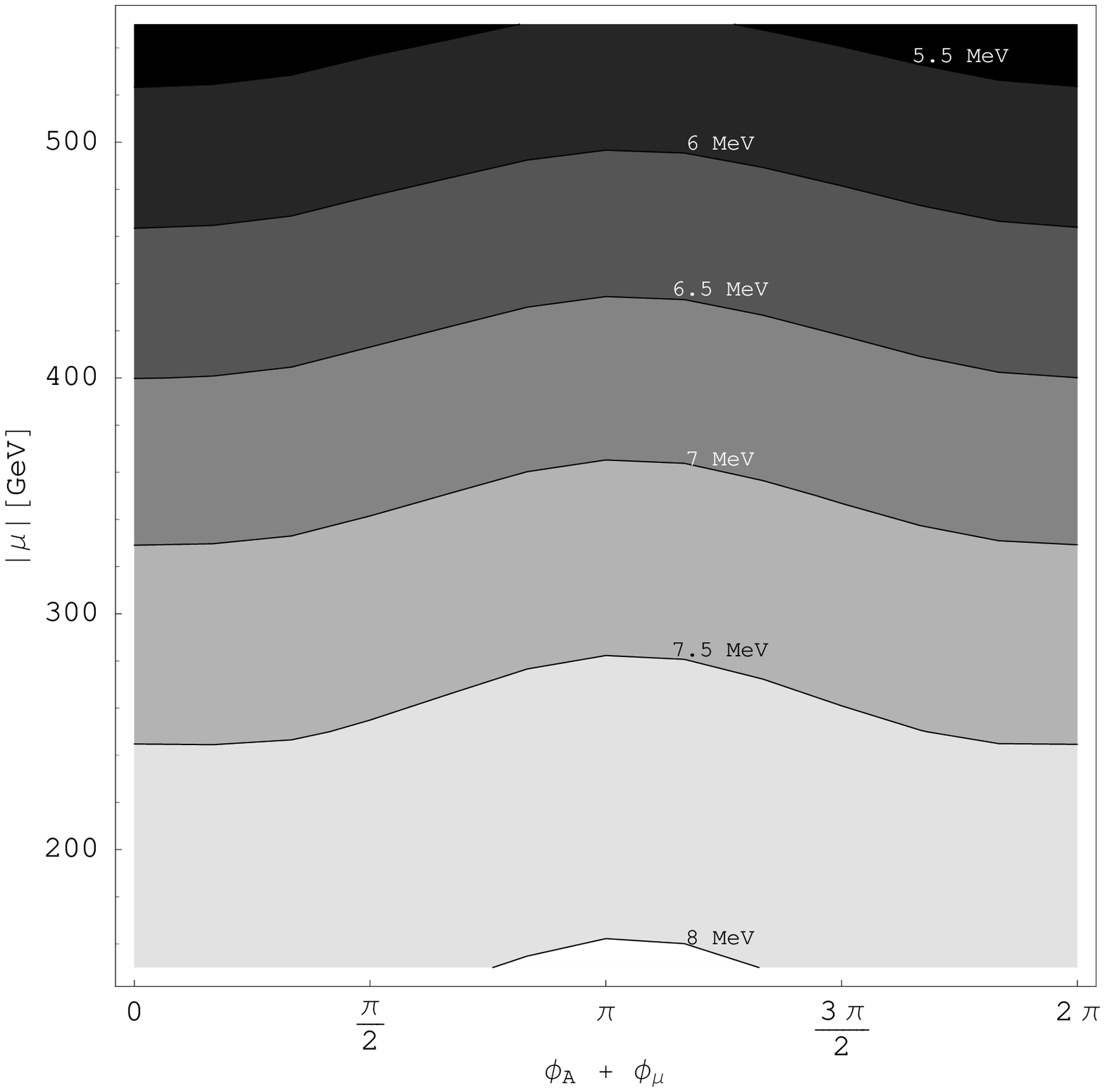}
\caption{Squark  contributions to $\de \MW$  as function of the phase
 $(\phi_A + \phi_\mu)$, where $\phi_A \equiv \phi_{\At} = \phi_{\Ab}$,
for different values of the common sfermion mass $\msusy=500$, $600$,
 $1000$ GeV.
The other relevant SUSY parameters are set to  $\tb = 5$, 
$|A_{t,b}|=2\msusy$, $|\mu|=900$ GeV.} 
\label{deltaMWSfermMUEPhase}
\end{center}
\end{figure}

\begin{figure}[htb!]
\begin{center}
\includegraphics[width=7.7cm,height=9.0cm]{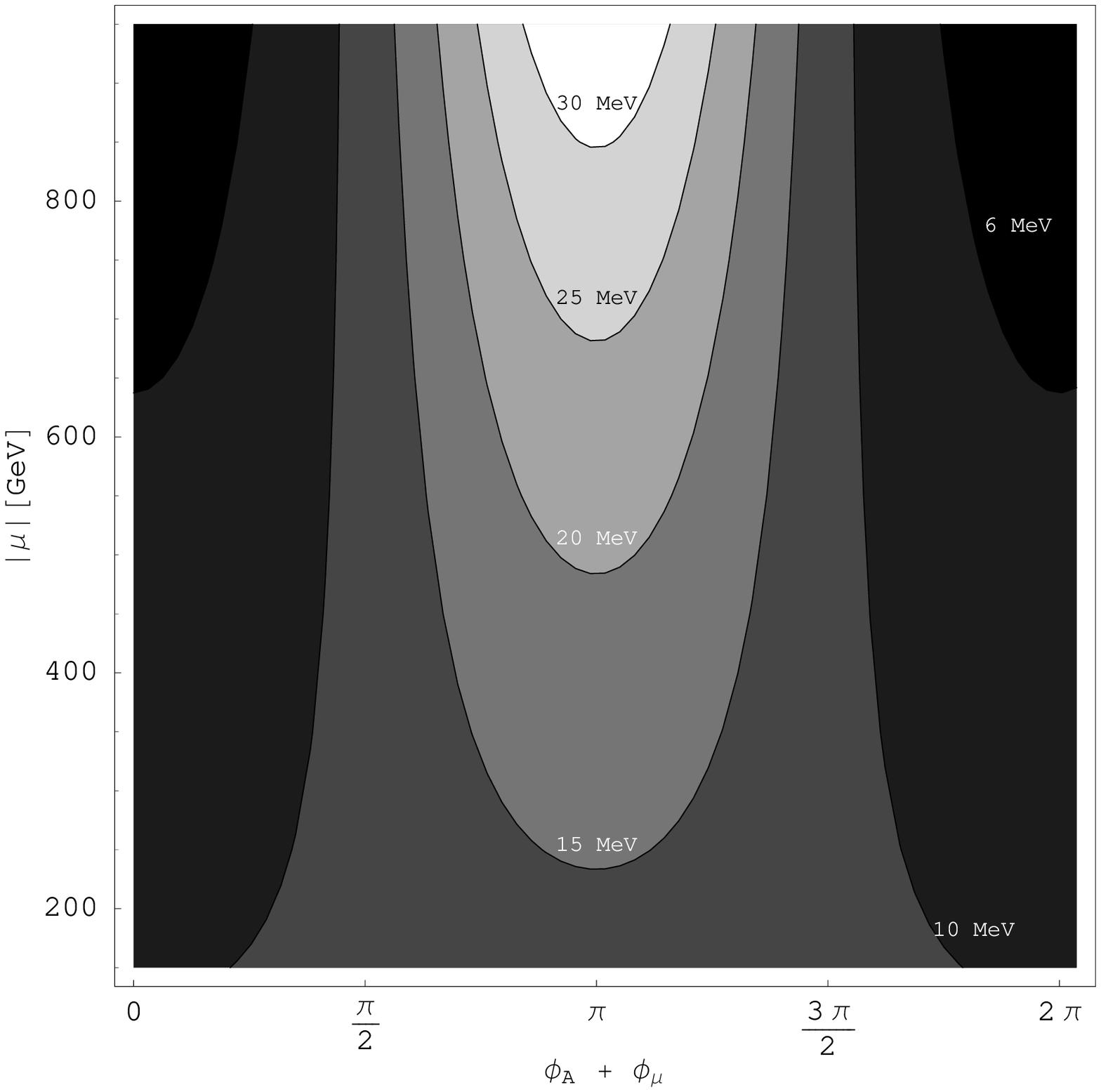}
\hspace{.2cm}
\includegraphics[width=7.7cm,height=9.0cm]{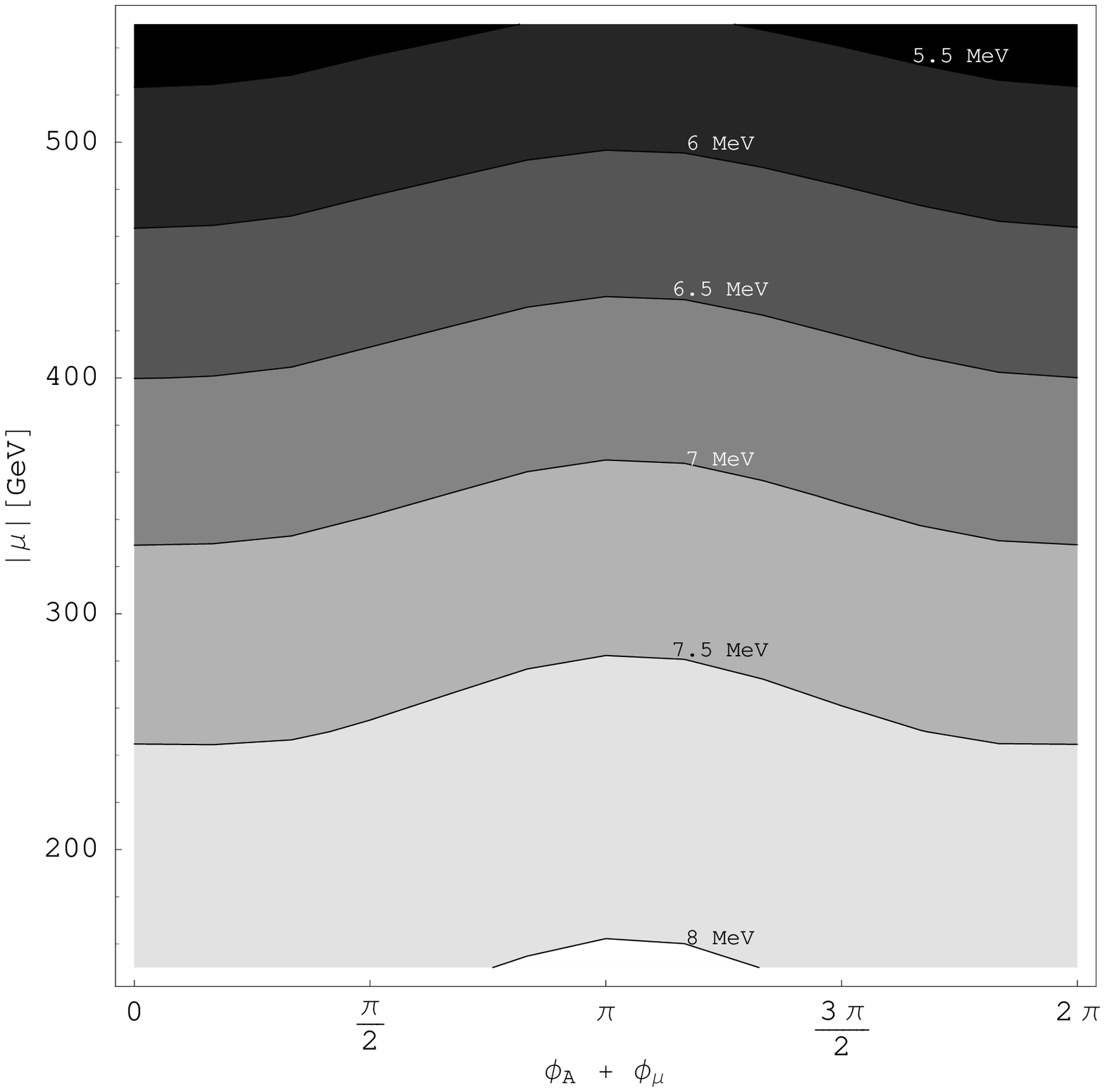}
\caption{Contour lines of the squark contributions to 
$\de\MW$ in the plane of $(\phi_A + \phi_\mu)$ 
and $|\mu|$, where $\phi_A \equiv \phi_{\At} = \phi_{\Ab}$.
The left plot shows a scenario with $\tb = 5$, $\msusy = 500 \gev$,
$|A_{t,b}|=1000 \gev$, while in the right plot $\tb = 30$, 
$\msusy = 600 \gev$, $|A_{t,b}|=1200 \gev$.}
\label{deltaMWMMUEphase}
\end{center}
\end{figure}

The phase dependence is illustrated in 
\reffis{deltaMWSfermMUEPhase} and \ref{deltaMWMMUEphase}, where the
squark loop contributions to $\de\MW$ (evaluated from \refeq{deltaMW})
are shown as function of the phase
combination $(\phi_A+\phi_\mu)$ with $\phi_{\At}=\phi_{\Ab}$. Since the
phases enter only via $|X_{t,b}|$, their influence is most significant
if all terms in eqs.~(\ref{XtRel}), (\ref{XbRel}) are of a similar
magnitude. This is the case if $\tb$ is rather small and $|\mu|$
and $|A_{t,b}|$ are of the same order. Such a situation is displayed
in \reffi{deltaMWSfermMUEPhase} and the left panel of 
\reffi{deltaMWMMUEphase}, where $\tb=5$ and $|\At|=|\Ab|=2\msusy$
has been chosen. \reffi{deltaMWSfermMUEPhase} shows the 
effect on $\de\MW$ from varying the phase $(\phi_A+\phi_\mu)$ for a
fixed value of $|\mu|=900 \gev$ and $\msusy=500$, $600$, $1000 \gev$,
while in \reffi{deltaMWMMUEphase} the squark sector contributions to
$\de\MW$ are shown as contour lines in the plane of $(\phi_A+\phi_\mu)$
and $|\mu|$. In the scenario with $\tb=5$ (\reffi{deltaMWSfermMUEPhase}
and left panel of \reffi{deltaMWMMUEphase}) the variation of the complex
phase $(\phi_A+\phi_\mu)$ can amount to a shift in the $W$~boson mass of
more than $20 \mev$. The most pronounced phase dependence is obtained
for the largest sfermion mixing, i.e.\ the smallest value
of $\msusy$ and the largest value of $|\mu|$.

The right panel of \reffi{deltaMWMMUEphase} shows a scenario where $\tb$
is rather large, $\tb=30$. As a consequence, $|X_t|\approx |\At|$ and
$|X_b|\approx|\mu\tb|$, so that the absolute values of $X_t$ and $X_b$
depend only very weakly on the complex
phases. The plot clearly displays the resulting much weaker phase
dependence compared to the scenario in the left panel of
\reffi{deltaMWMMUEphase}.
The variation of the complex phase gives rise
only to shifts in $\MW$ of less than $0.5$~MeV, while changing $|\mu|$
between $100$ and $500$~GeV leads to a shift in $\MW$ of about 2~MeV.

\begin{figure}[htb!]
\begin{center}
\includegraphics[width=10cm,height=8.4cm]{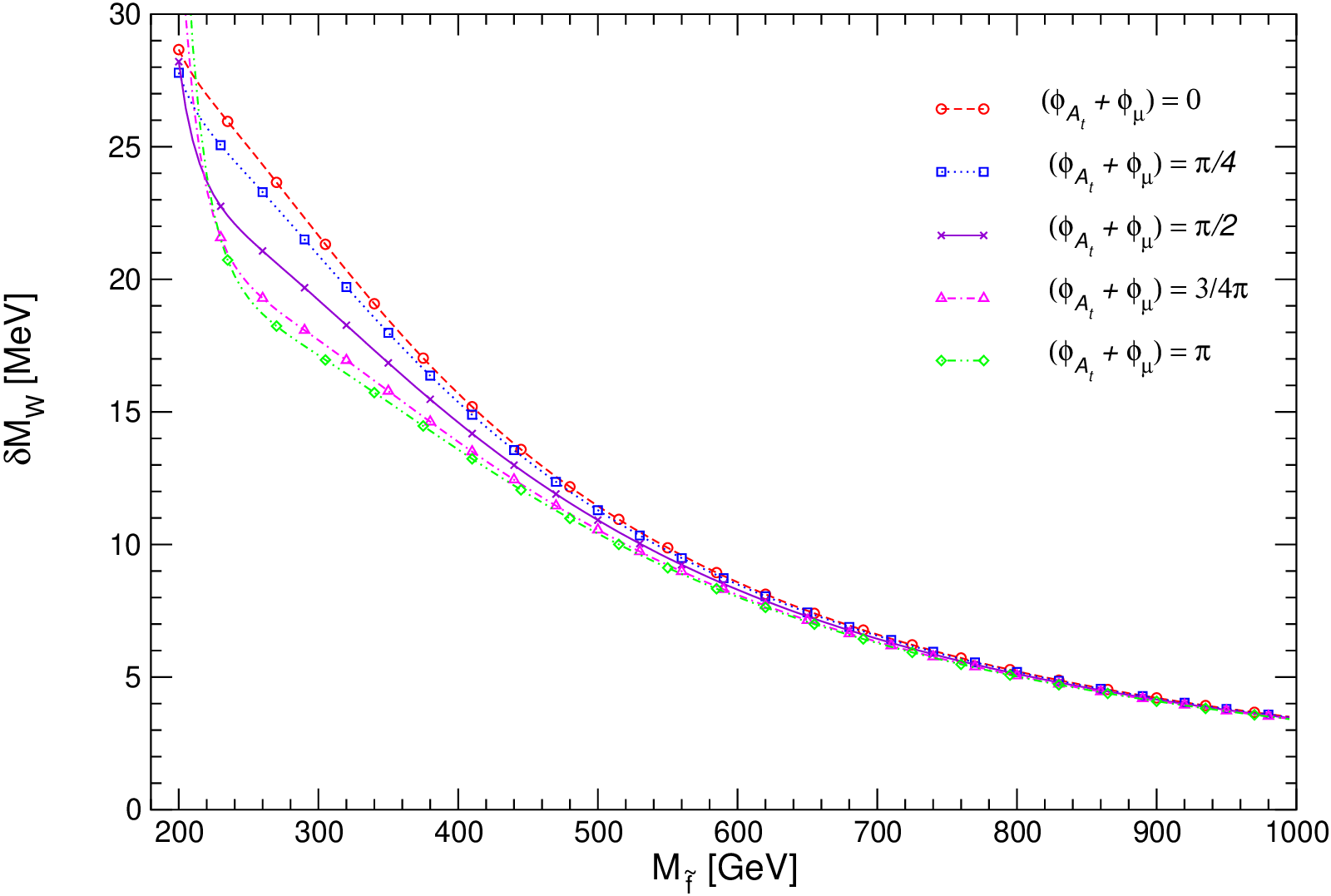}
\caption{Squark contributions to $\de \MW$  as function of a common
  sfermion mass $\msusy$ for different values of 
  $(\phi_{\At} + \phi_\mu)$. The
  other relevant SUSY parameters are: $|A_{t,b}|=350 \gev$,
  $(\phi_{\Ab} + \phi_\mu) =0$, $|\mu| =300\gev$ and $\tb = 10$.} 
\label{deltaMWMSUSYAt}
\end{center}
\end{figure}

\begin{figure}[htb!]
\begin{center}
\includegraphics[width=10cm,height=8.4cm]{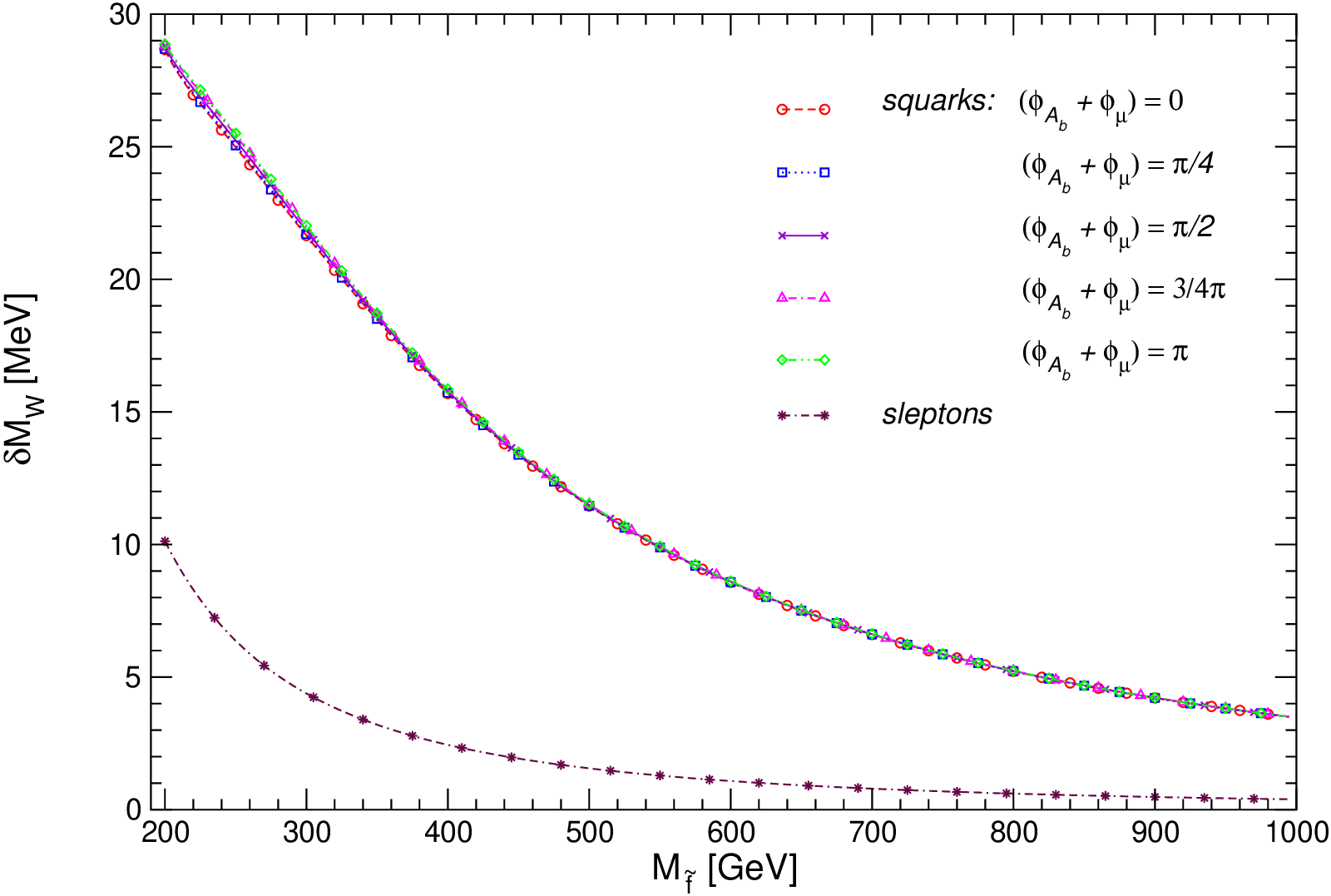}
\caption{Squark and slepton contributions to $\de \MW$  as function
  of a common sfermion mass $\msusy$ for different values of
  $(\phi_{\Ab} + \phi_\mu)$. 
  The other relevant SUSY parameters are: $|A_{t,b}|=350 \gev$, 
  $(\phi_{\At} + \phi_\mu) =0$, $|\mu| =300\gev$ and $\tb = 10$.} 
\label{deltaMWMSUSYAb}
\end{center}
\end{figure}

In the following plots we discuss the dependence of the sfermion loop
contributions to $\de \MW$ on the common sfermion mass $\msusy$.
In \reffi{deltaMWMSUSYAt} we show the squark contributions 
as function of $\msusy$ for various values of $(\phi_{\At} +
\phi_\mu)$ and for $(\phi_{\Ab}+\phi_\mu) =0$.
The intermediate value $\tb=10$ is chosen, and
$|A_{t,b}| = 350 \gev$, $|\mu| = 300 \gev$.
In agreement with the discussion above, the phase dependence is 
relatively small.
It leads to a shift of about $5$~MeV in $\MW$ for low values of
$\msusy\approx250$~GeV, where the stop mixing is large.
The total squark contributions can shift the prediction of
$\MW$ by up to 30 MeV for small $\msusy$. For large $\msusy$ the squark
contributions show the expected decoupling behaviour.
However, even for sfermion masses as large as $\msusy=1000$~GeV the 
shift in $\MW$ is
still about $4$~MeV, i.e.\ half the size of the
anticipated GigaZ accuracy.

In \reffi{deltaMWMSUSYAb} we show the squark and slepton contributions
for the same parameters as before except that
$(\phi_{\At}+\phi_\mu) =0$ and $(\phi_{\Ab} + \phi_\mu)$ is varied.
The effect of the phase $(\phi_{\Ab} + \phi_\mu)$ in the squark
contributions is negligible since the sbottom mixing is small and 
moreover $|X_b|\approx|\mu|\tb$, making its phase dependence 
insignificant. 
The slepton contributions (entering via the diagrams in
\reffi{fig:SquarkSleptoneloop}) yield a shift in $\MW$ of up to 10~MeV
for small $\msusy$, i.e.\ about a third of the squark contributions.
Even for the slepton contributions the dominant effect (about 60\% of
the total shift in $\MW$) can be associated with $\De\rho$, as a
consequence of the D-term splitting of the sleptons.
For large $\msusy$ the slepton contributions show
the expected decoupling behaviour.


\subsubsection{Chargino and neutralino sector dependence}
\label{subsec:cndep}

In this subsection we analyse the one-loop contributions from
neutralinos and charginos to $\de \MW$, entering
via the self-energy, vertex and box diagrams shown in 
\reffis{fig:NCselfoneloop} and \ref{fig:NCvertexboxoneloop}. 
In this sector the parameters $M_1$,
$M_2$ and $\mu$ can be complex. However, there are only two physical
complex phases since one of the two phases of $M_1$ and $M_2$ can be
rotated away.
As commonly done we choose to rotate away the phase of $M_2$.
Generally, the phase dependence in this sector can be expected to be smaller
than in the sfermion sector since the chargino/neutralino mass matrices are
dominated by their diagonal elements, and the mass eigenvalues are mainly
determined by $|\mu|$, $|M_{1,2}|$, so that their phase dependence is
small.

\begin{figure}[htb!]
\begin{center}
\includegraphics[width=7.7cm,height=8.4cm]{DeltaMWMUE.eps}
\hspace{.2em}
\includegraphics[width=7.7cm,height=8.4cm]{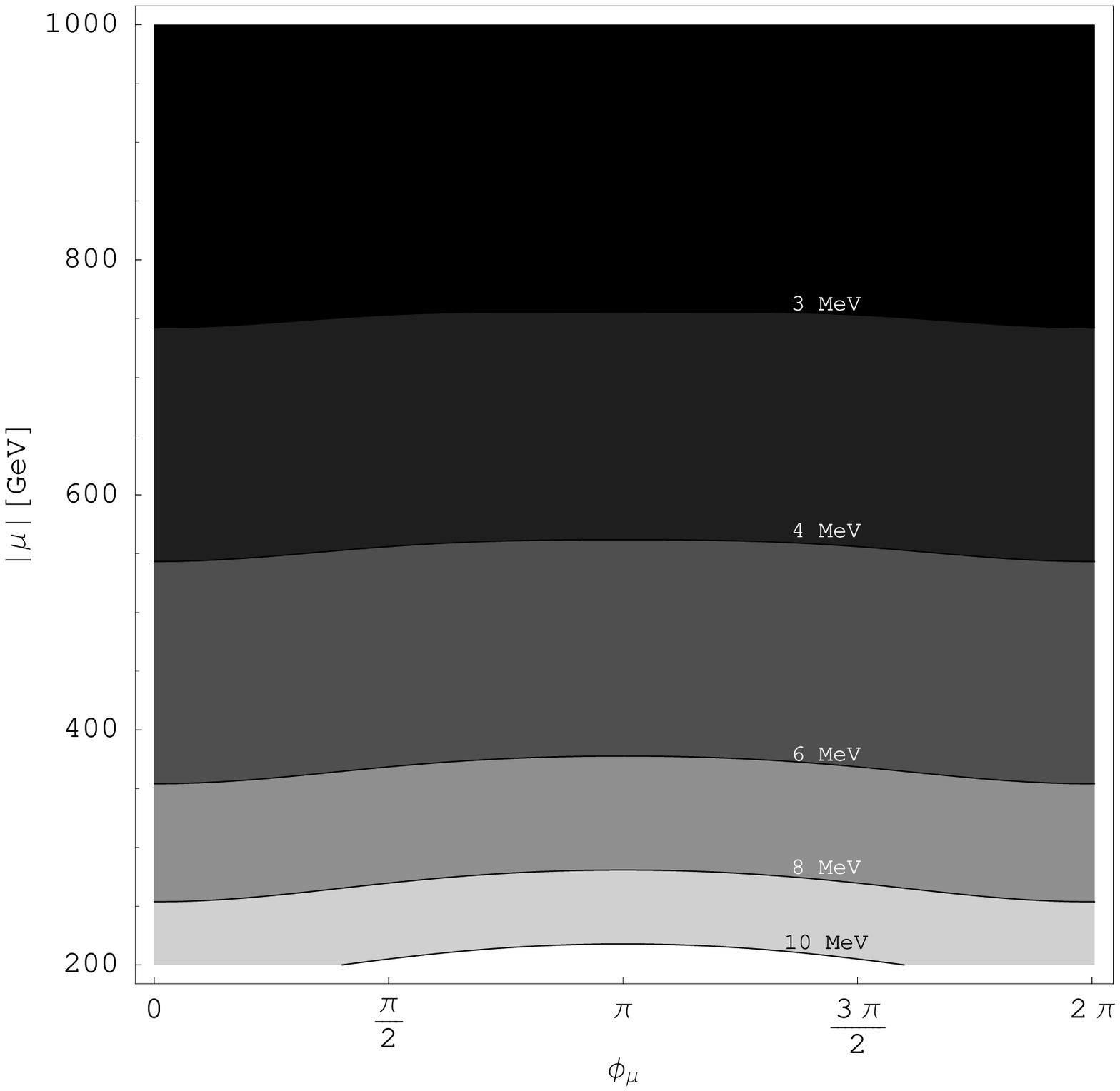}
\caption{Neutralino and chargino contributions to $\de \MW$. The left
plot shows $\de\MW$ as function of $|\mu|$ for different values of 
$\phi_{\mu}$, while the right plot 
displays the contour lines of $\de\MW$ in the
$\phi_\mu$--$|\mu|$ plane. The other relevant SUSY
  parameters are: $\phi_{M_1} = \phi_{M_2} = 0$,
  $|M_1| = |M_2|  = 200\gev$,
  $\tb = 10$ and $\msusy = 500\gev$.} 
\label{deltaMWMUE}
\end{center}
\end{figure}

The phase dependence is illustrated by 
 \reffi{deltaMWMUE}, where the chargino/neutralino contributions to
 $\MW$ are shown as function of $|\mu|$ (left panel) and as contour
lines in the $\phi_\mu$--$|\mu|$ plane (right panel)
 for $\phi_{M_1}=0$. The other parameters are $\msusy = 500 \gev$,  
$M_1 = M_2 = 200 \gev$ and $\tb = 10$. The effect of varying
 $\phi_\mu$ is much smaller than the overall contribution of the
 chargino/neutralino sector. In the scenario of  
\reffi{deltaMWMUE} the chargino/neutralino contributions lead to a shift
in the prediction for $\MW$ of up to $11 \mev$ for small $|\mu|$, while
the effect of varying $\phi_\mu$ does not exceed $1 \mev$.
We have checked that also the dependence on $\phi_{M_1}$ is
insignificant.

\begin{figure}[htb!]
\begin{center}
\includegraphics[width=10cm]{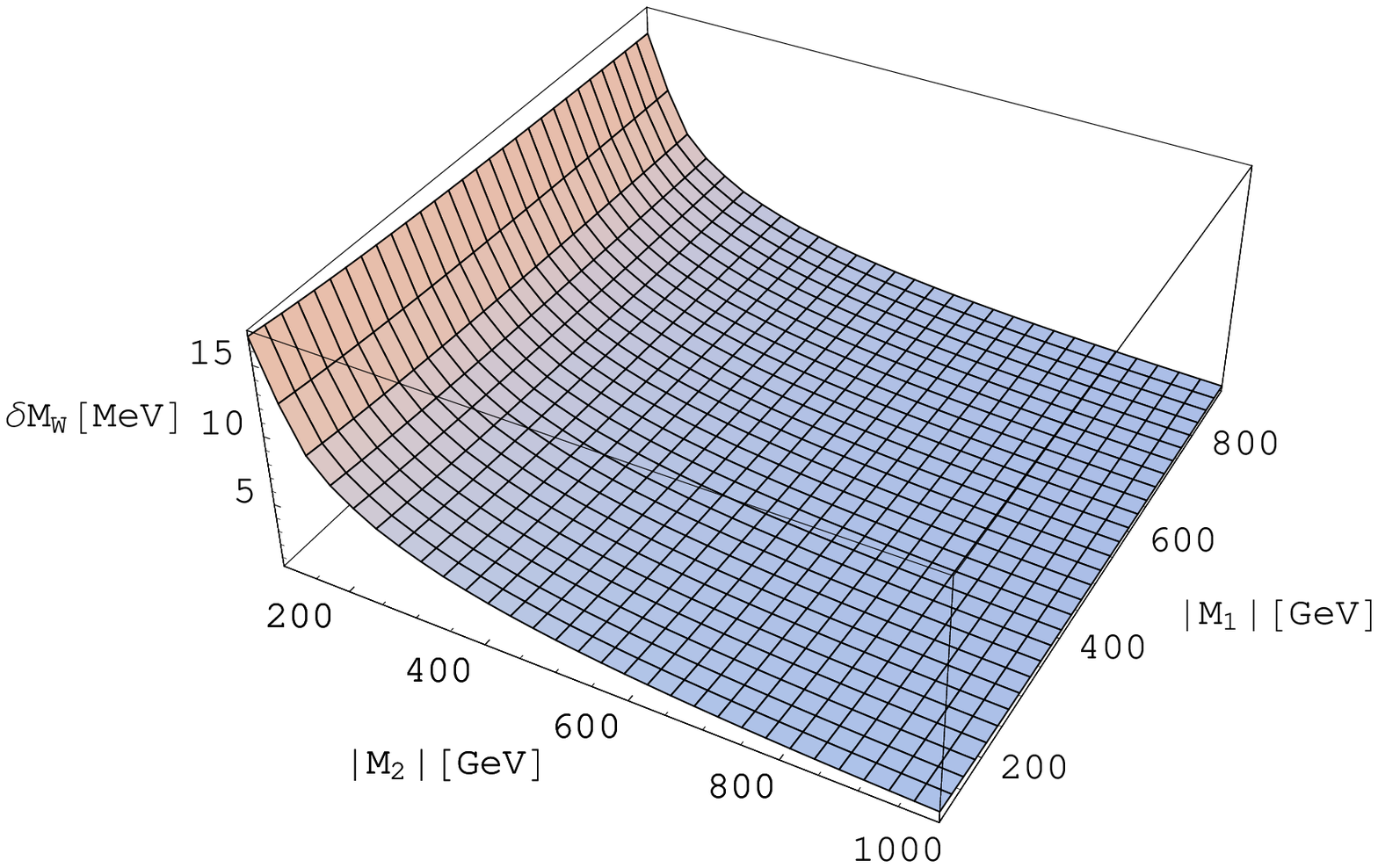}
\caption{Neutralino and chargino contributions to $\de \MW$ as function
  of $|M_1|$ and $|M_2|$.  The other relevant SUSY
  parameters are: $\phi_{M_1}=\phi_{M_2}=\phi_{\mu}=0$, $|\mu| = 300 \gev$,
  $\tb = 10$ and $\msusy = 250 \gev$.} 
\label{deltaMWM1M23D}
\end{center}
\end{figure}

In \reffi{deltaMWM1M23D} we investigate the impact of varying 
$|M_1|$ and $|M_2|$ for zero complex phases.
The other relevant parameters are
$\msusy = 250 \gev$, $|\mu| = 300 \gev$, $\tb = 10$.
The shift in $\MW$ induced by varying
$|M_2|$ can reach up to $15 \mev$ (i.e.\ the anticipated
LHC precision). This is larger than the maximum shift in
\reffi{deltaMWMUE} because of the smaller sfermion masses. 
On the other hand, the effect of varying $|M_1|$ stays below $\sim 2
\mev$.

\subsubsection{Gauge boson and Higgs sector dependence}
\label{subsec:Higgs}

Finally we discuss the one-loop effects on $\MW$
from the gauge and Higgs
sectors. In order to identify the genuine SUSY
effects, we compare the contribution in the MSSM with the one in the SM
(where the Higgs-boson mass in the SM is set equal to
the mass of the light $\cp$-even Higgs boson of the MSSM, 
$\MHSM = \Mh^{\rm MSSM}$). The corresponding diagrams are
shown in \reffis{fig:GHMSSMSMlikeselfoneloop},
\ref{fig:GHMSSMSUSYselfoneloop} and \ref{fig:GHMSSMboxvertexoneloop},
where only the diagrams shown in \reffi{fig:GHMSSMSUSYselfoneloop}
differ in the MSSM and the SM. The parameters governing the Higgs sector
at the tree level are 
$M_A$ and $\tb$ in the MSSM and $\MHSM$ in the
SM. Since we incorporate higher-order contributions into the predictions
for the MSSM Higgs masses and mixing angles, which are evaluated using
the program \fh, further SUSY parameters enter the prediction for the
gauge-boson and Higgs sector contribution. The effect of complex
phases entering via the MSSM Higgs sector is formally of two-loop order. 
We therefore restrict to real parameters in this subsection.

\begin{figure}[htb!]
\begin{center}
\includegraphics[width=13cm]{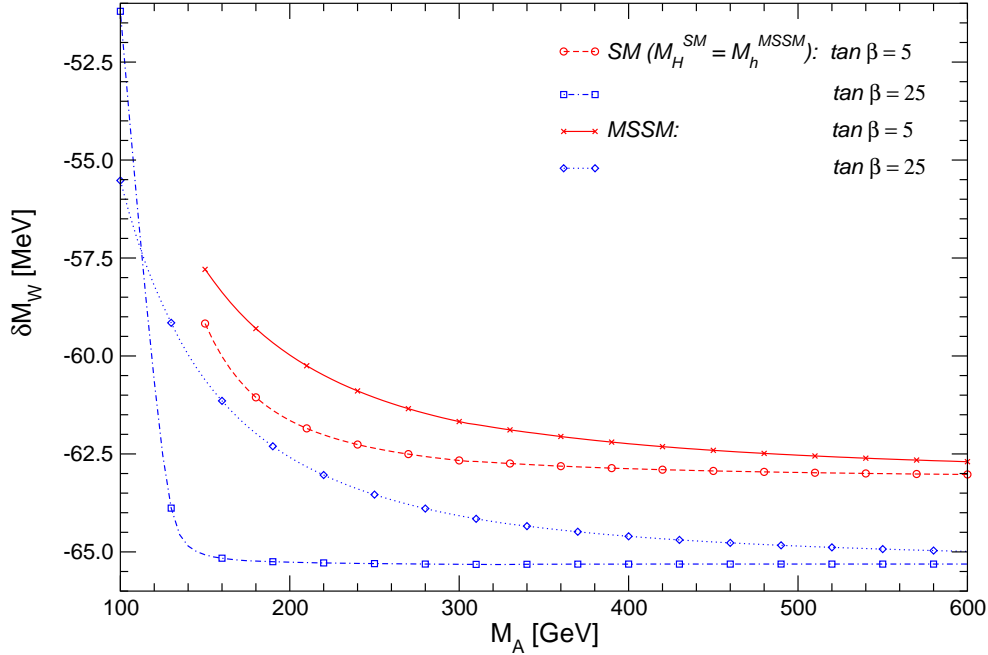}
\caption{The contribution $\de\MW$ to the prediction for the $W$~boson
mass from the gauge-boson and Higgs sector contributions in the MSSM and 
the SM, where the SM contributions are evaluated for 
$\MHSM=\Mh^{\rm MSSM}$. The shift $\de \MW$ is given as a function of
the $\cp$-odd Higgs-boson mass $\MA$ and $\tb = 5, 25$.
The other SUSY parameters were chosen to be:
  $\msusy=600\gev$, $A_{t,b}=1200\gev$,  $\mu =500\gev$,
  $\mgl=500 \gev$, $M_2 = 500 \gev$.}  
\label{deltaMWGHFig}
\end{center}
\end{figure}

In \reffi{deltaMWGHFig} the shift 
$\de\MW$ is given as a function of $\MA$ in the
MSSM and in the SM (with $\MHSM = \Mh^{\rm MSSM}$).
The parameter $\tb$ is fixed to $\tb = 5,25$,
which affects $\Mh^{\rm MSSM}$ and accordingly also $\MHSM$.
The other SUSY
parameters are chosen as $\msusy = 600 \gev$, $A_{t,b}=1200\gev$,
$\mu=500\gev$, $\mgl = 500 \gev$, $M_2=500\gev$.
\reffi{deltaMWGHFig}  shows
that the overall effect of the gauge and Higgs boson sector is rather
large,  up to $- 60 \mev$, which corresponds to about twice the current 
experimental error on $\MW$. 
The shift in the $W$~boson mass obtained in the MSSM is slightly larger
than in the SM.
However, this genuine SUSY effect, i.e.\ the difference between
the MSSM and the SM value of $\de\MW$, is always below 
$\sim 2 \mev$ for $\tb = 5$ and smaller than $\sim 5 \mev$ for 
$\tb = 25$.
The genuine SUSY effect is therefore 
below the anticipated GigaZ accuracy.

\subsection{Prediction for \boldmath{$\MW$}}
\label{subsec:MWpred}

In this section we
combine all known contributions as described in \refse{sec:higherorders}
and analyse the
parameter dependence of this currently most accurate MSSM prediction for $\MW$
in various scenarios. These predictions are
compared with the current experimental value for
$\MW$~\cite{LEPEWWG2}, 
\begin{equation}
\MW^{\rm exp} = 80.404 \pm 0.030 \gev~.
\label{MWexp}
\end{equation}


\subsubsection{Dependence on SUSY parameters}
\label{subsubsec:SUSYpara}

\begin{figure}[htb!]
\begin{center}
\includegraphics[width=13cm,height=8.0cm]{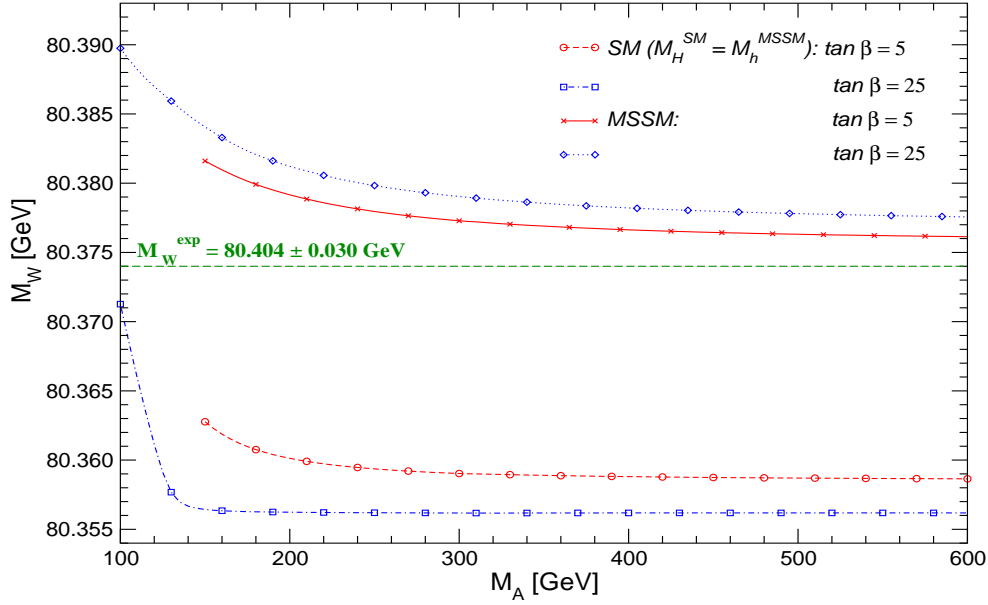}
\caption{Prediction for $\MW$ in the MSSM and the SM, where the SM
contributions are evaluated for $\MHSM=\Mh^{\rm MSSM}$. The prediction
for $\MW$ is shown as function of the $\cp$-odd Higgs-boson mass
$\MA$ for $\tb = 5,  25 $. 
The other SUSY parameters are:
$\msusy=600\gev$, $A_{t,b}=2 \, \msusy$,  $\mu =500\gev$, $\mgl=500 \gev$, 
$M_2 = 500 \gev$.}  
\label{MWGHFig}
\end{center}
\end{figure}

We start by comparing our full MSSM prediction for $\MW$ with the
corresponding SM value (with $\MHSM = \Mh$) as a function of
$\MA$ in \reffi{MWGHFig}. Like in \reffi{deltaMWGHFig} 
the other parameters are
$\msusy=600\gev$, $A_{t,b}=1200\gev$,  $\mu =500\gev$, 
$\mgl=500\gev$, $M_2 = 500 \gev$. 
$\tb$ is set to $\tb = 5, 25$. It can be seen in
\reffi{MWGHFig} that for this set of parameters the MSSM prediction is
about $20 \mev$ higher than the SM prediction. While the MSSM
prediction is within $1 \si$ of the experimental value of $\MW$, 
the SM prediction lies in the 1--$2 \si$ interval. 

\begin{figure}[htb!]
\begin{center}
\includegraphics[width=13cm,height=8.0cm]{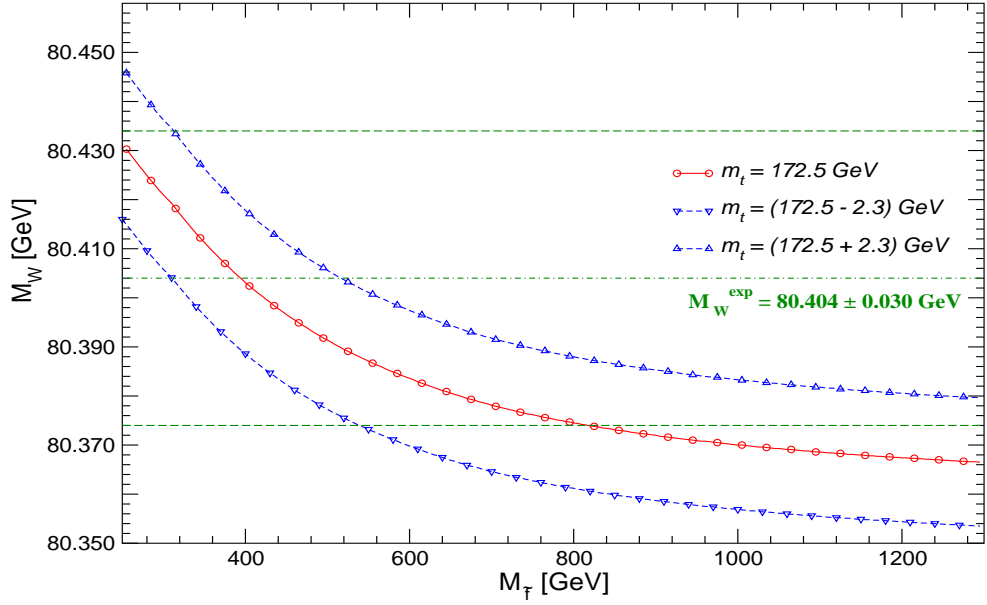}
\caption{Prediction for $\MW$ as function of a common sfermion mass for
$\mt = (172.5 \pm 2.3) \gev$. The other SUSY parameters are set to be: 
$A_{t,b}=2 \,\msusy$, $\tb = 10$, $\mu =300\gev$, $\mgl=300 \gev$,
  $\MA=300\gev$, $M_2 = 300 \gev$.}
\label{MWMSf}
\end{center}
\end{figure}

In \reffi{MWMSf} we show the prediction for $\MW$ as a function of
$\msusy$ and indicate how this prediction changes if the top-quark mass
is varied within its experimental $1\si$ interval,
$\mt = (172.5 \pm 2.3) \gev$~\cite{newestmt}. 
The other parameters are $A_{t,b} = 2 \, \msusy$, 
$\mu = \MA = \mgl = M_2 = 300 \gev$, and $\tb = 10$.
The result is compared
with the current experimental value of $\MW$. It can be seen in
\reffi{MWMSf} that 
the observable $\MW$ exhibits a slight preference for a relatively
low SUSY scale, see also \citeres{ehow3,ehow4} for a recent discussion
of this issue. For the current experimental central value of the
top-quark mass, $\MW$ lies in the experimental $1\si$-interval only
for a SUSY mass scale of $\msusy \lsim 800 \gev$.
Increasing $\mt$ by one standard deviation allows 
$\msusy$ up to at least $1300\gev$ at the $1\si$ level for this set of
SUSY parameters. 

\begin{figure}[htb!]
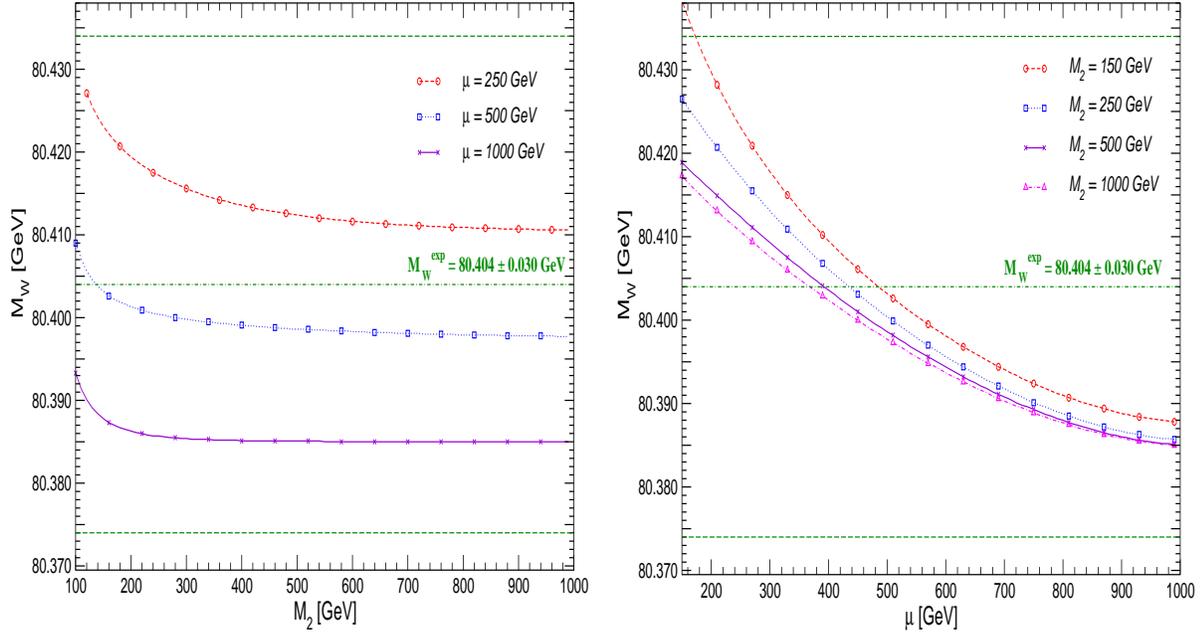

\begin{center}
\includegraphics[width=7.7cm,height=8.4cm]{MWM2.eps}
\hspace{.2em}
\includegraphics[width=7.7cm,height=8.4cm]{MWMUE.eps}
\caption{Prediction for $\MW$ as function of 
$M_2$ for $\mu = 250\gev, \,  500 \gev, \, 1000\gev $ (left plot) and
as function of $\mu$ for $M_2 = 150\gev, \, 250\gev, \, 500 \gev, \, 
1000\gev $ (right plot). The other SUSY
  parameters are: $\msusy = 300\gev$, $A_{t,b}=2 \, \msusy$, $\tb = 10$,
  $\mgl=600 \gev$, $\MA=1000\gev$.}
\label{MWM2MUE}
\end{center}
\end{figure}

In \reffi{MWM2MUE} we show the dependence of $\MW$ on
$\mu$ and $M_2$. The other parameters are set to $\msusy = 300 \gev$,
$A_{t,b} = 2\,\msusy$, $\MA = 1000 \gev$, $\mgl = 600 \gev$, 
$\tb = 10$.
As can be seen in \reffi{MWM2MUE}, varying $\mu$ between
about $200 \gev$ and $1000 \gev$ results in a downward shift of more 
than $40 \mev$ in $\MW$. 
This strong dependence on the $\mu$ parameter is due to the neutralino
and chargino as well as the squark contributions (one- and two-loop)
to $\Delta r$. The sensitivity to the $\mu$ parameter from both MSSM
particle sectors adds up and leads to the large shifts shown in
\reffi{MWM2MUE}. The neutralino and chargino contributions are
responsible for roughly one third of the $40 \mev$ shift for small
$M_2$ and become negligible  for large $M_2$, with the remaining MSSM
parameters specified as given above. 
 The shift induced by varying $M_2$ between 
about $100 \gev$ and $1000 \gev$ (as explained above, $M_2$ and $M_1$
are varied simultaneously according to \refeq{eq:GUT}) amounts up to
about $15 \mev$ in $\MW$.
For the relatively small value chosen for the common sfermion mass,
$\msusy=300\gev$,
all combinations of $\mu$ and $M_2$ yield a result within $1 \si$
of the experimental result of $\MW$.
Using instead a larger SUSY mass scale of, for instance,
$\msusy = 600 \gev$ would result in $\MW$
values within the experimental $1 \si$ interval only for 
$\mu \lsim 500 \gev$.

\begin{figure}[htb!]
\vspace{-3em}
\begin{center}
\includegraphics[width=13cm,height=8cm]{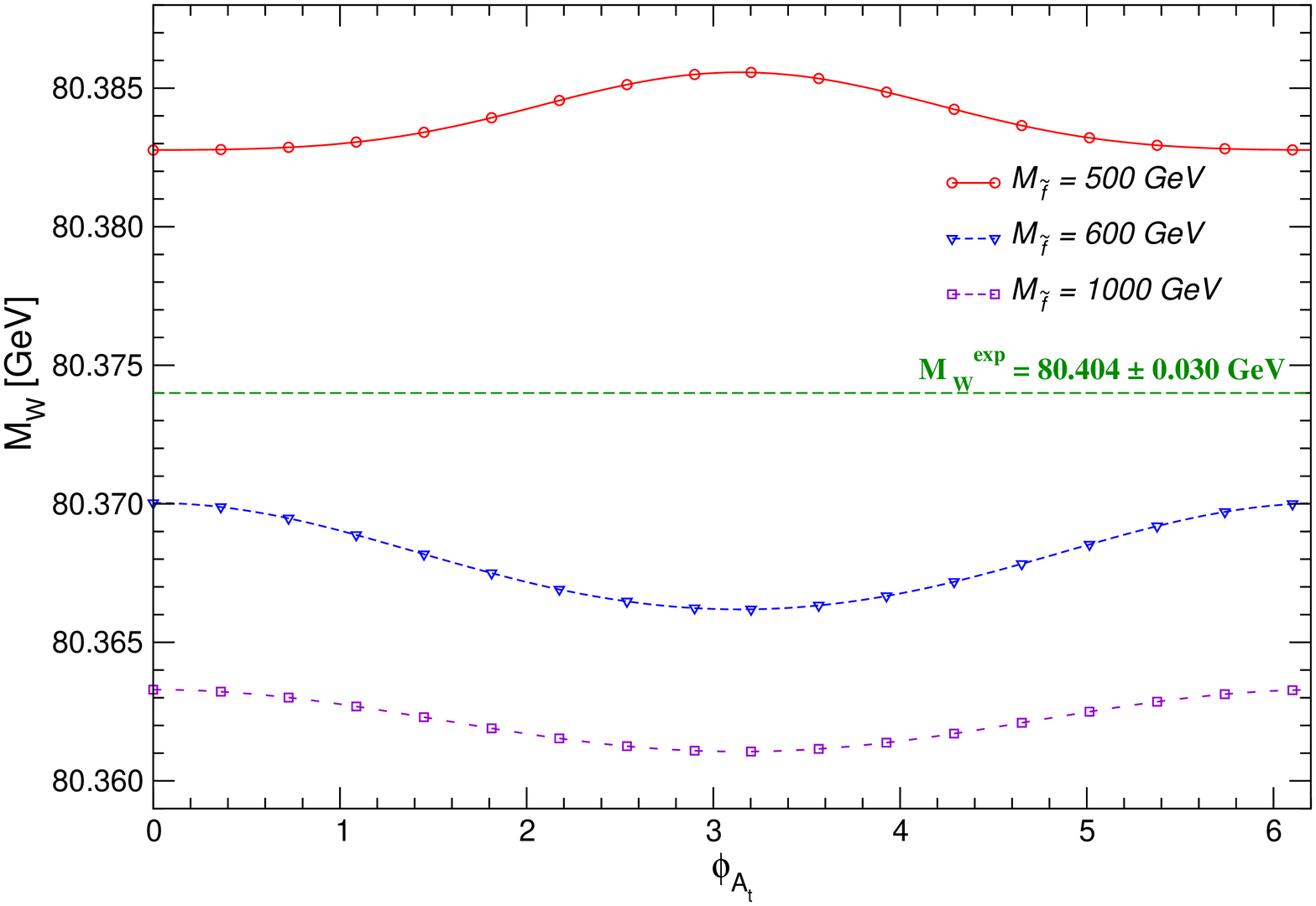}
\caption{Prediction for $\MW$ as a function of the phase of $\At$ 
using relation (\ref{phases2L}). The SUSY parameters are $|A_{t,b}|=1000 \gev$,
$\phi_{\Ab} = 0$, $\tb = 10$, $\mgl=500 \gev$, $\MA=500\gev$, 
$M_2 = 250\gev$ and $\mu = 500 \gev$. } 
\label{AtphaseMW2L}
\end{center}
\end{figure}

Finally we discuss the effect of varying the complex phase $\phi_{\At}$
on the prediction for $\MW$. As explained above, we use our complete
one-loop result for the phase dependence and employ \refeq{phases2L}
to approximate the effect of the complex phases at the two-loop level.
In \reffi{AtphaseMW2L} the prediction for $\MW$ is shown as a function
of $\phi_{\At}$ for $|A_{t,b}|=1000 \gev$, $\phi_{\Ab} = 0$, 
$\tb = 10$, $\mgl=500 \gev$, $\MA=500\gev$, $M_2 = 250\gev$ and 
$\mu = 500 \gev$. The results are plotted for $\msusy = 500\gev, 600 \gev$
and $1000 \gev$. The dependence on $\phi_{\At}$ is at most of the
order $2 \mev$ for $\msusy = 500\gev, 600 \gev$. For heavier sfermions
with $\msusy = 1000\gev$ only a $1 \mev$ shift in $\MW$ can be
observed. 

As explained above, our result for $\MW$ goes beyond the results
previously known in the literature~\cite{PomssmRep} because of the
inclusion of complex phases and an improved treatment of higher-order
SM contributions. We have checked that for real parameters our new
result agrees with the previously most advanced
implementation~\cite{PomssmRep} typically within about $5 \mev$.


\subsubsection{The SPS inspired benchmark scenarios}
\label{subsubsec:sps}

In this subsection we show $\MW$ in the SPS~1a, SPS~1b and SPS~5
benchmark scenarios~\cite{sps}. This should give an indication of the $\MW$
prediction within ``typical'' constrained MSSM (CMSSM) scenarios. 
In the original definition the SPS
parameters are \drbar\ parameters. Here we treat them as on-shell
input parameters for simplicity, since the effects of the \drbar\ to on-shell
transition are expected to be small and therefore irrelevant for the
qualitative features that we discuss. In order to
analyse the dependence of $\MW$ on the scale of supersymmetry we
introduce a scale factor. Every SUSY parameter of mass dimension of
the considered SPS point is multiplied by this parameter, i.e.\
$\MA=(\textup{scalefactor})\times \MA^{\textup{SPS}}$, 
$M_{{\tilde F},{\tilde F'}}=(\textup{scalefactor})\times 
                            M_{{\tilde F},{\tilde F'}}^{\textup{SPS}}$,
$A_{t,b}=(\textup{scalefactor})\times A_{t,b}^{\textup{SPS}}$,
$\mu=(\textup{scalefactor})\times\mu^{\textup{SPS}}$,
$M_{1,2,3}=(\textup{scalefactor})\times M_{1,2,3}^{\textup{SPS}}$.

\begin{figure}[htb!]
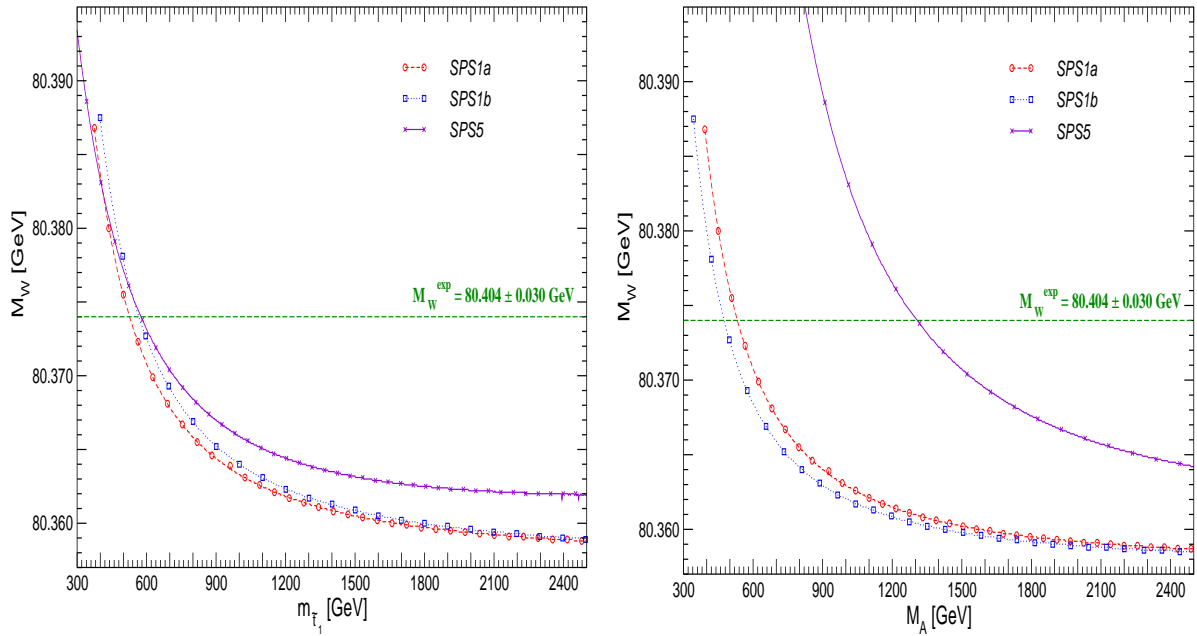

\begin{center}
\includegraphics[width=7.7cm,height=8.4cm]{SPSMSf133.eps}
\hspace{.2em}
\includegraphics[width=7.7cm,height=8.4cm]{SPSMA0.eps}
\caption{Prediction for $\MW$ within the SPS~1a, SPS~1b and SPS~5 inspired
  scenarios. $\MW$ is shown as a function of $\mste$, the
  lighter of the two stop squarks (left plot), and as a function of
  $\MA$, the mass of the $\cp$-odd Higgs boson (right plot). The SPS 
  parameters of mass dimension are varied with the
  scale of supersymmetry as described in the text.} 
\label{SPSMSf133} 
\end{center}
\end{figure}

In \reffi{SPSMSf133} we show the result for the three SPS scenarios as
a function of the lighter $\Stop$ mass, $\mste$ (left plot), and as a
function of $\MA$ (right plot). The prediction for
$\MW$ is similar in all three scenarios. The variation of the 
dimensionful SUSY parameters shifts
the MSSM prediction for $\MW$ by up to $35 \mev$.
As can be seen in the left plot of \reffi{SPSMSf133},
agreement at the $1 \si$ level 
with the experimental result is obtained for $\mste \lsim 600 \gev$. 
Since we scale all dimensionful parameters simultaneously, 
the variation from small to large $\mste$ (left plot) is the same 
as the one from small to large $\MA$ (right plot).
However, for the same $\MA$ value the three
scenarios can differ also by up to $\sim 30 \mev$.

\begin{figure}[htb!]
\begin{center}
\includegraphics[width=13cm,height=8.0cm]{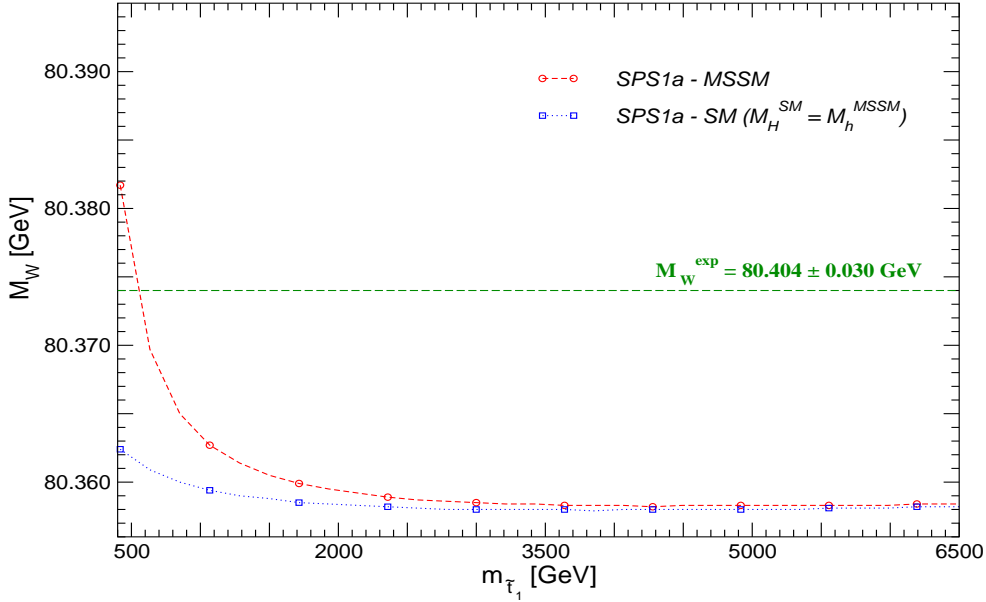}
\caption{Prediction for $\MW$ in the decoupling limit compared to the SM
prediction with $\MHSM = \Mh^{\rm MSSM}$. $\MW$ is shown as function of
the light stop mass $\mste$ in the SPS~1a benchmark scenario.
All dimensionful SUSY parameters are scaled with a common factor, as
described in the text.}
\label{SPSdecoupling}
\end{center}
\end{figure}

For large values of the SUSY mass scale one expects a decoupling
behaviour of the SUSY contributions, i.e.\ one expects that the
prediction for $\MW$ in the MSSM coincides with the SM prediction 
(for $\MHSM = \Mh^{\rm MSSM}$) in the limit of large SUSY masses. 
We analyse the decoupling behaviour in \reffi{SPSdecoupling} for the 
SPS~1a scenario. We compare the MSSM prediction with the corresponding SM
prediction of $\MW$ with $\MHSM = \Mh^{\rm MSSM}$. For 
$\mste \lsim 500 \gev$ large deviations between the MSSM and the SM
prediction can be observed. For $\mste \gsim 2500 \gev$ the
difference drops below the level of $1 \mev$, and for even larger
$\mste$ values the MSSM result converges to the SM result. It should be
noted in this context that the prescription described in 
\refse{subsec:SMandMSSM} (see \refeq{eq:obsSMSUSY}) has been crucial in
order to recover the most up-to-date SM prediction for $\MW$ in the decoupling
limit.


\subsubsection{The focus point scenario}
\label{subsubsec:focuspoint}

As we have seen in the previous section, a relatively light SUSY scale 
leads to a prediction for $\MW$ within the CMSSM (and of course also
the unconstrained MSSM) that is in slightly better agreement with the
experimental value of $\MW$ than the SM prediction. A region of the
CMSSM that has found a considerable interest in the last years is the
so-called focus point region~\cite{focus}. This region is characterized 
by a relatively small fermionic mass parameter, $m_{1/2}$, while 
the common scalar mass parameter $m_0$ is very large, and also $\tb$ is
relatively large.%
\footnote{The CMSSM is characterized in terms of three GUT-scale
parameters, the common fermionic mass parameter $m_{1/2}$, the common
scalar mass parameter $m_0$, and the common trilinear coupling $A_0$.
These high-scale parameters are supplemented by the low-scale parameter
$\tb$ and the sign of $\mu$.}
We now investigate whether it is also possible to obtain a prediction
for $\MW$ that is in better agreement with the experimental value than
the SM prediction if the MSSM parameters are restricted to the focus
point region.

We have evaluated three representative scenarios, using 
$\mt = 172.5 \gev$, $\tb = 50$ and $\mu > 0$. We have chosen a point
with the currently lowest value of $m_{1/2}$ in the focus point region 
for which the dark matter density is allowed  
by WMAP and other cosmological data (see e.g.\ \citere{ehow4}
for a more detailed discussion) and two further points with higher
$m_{1/2}$ (and higher $m_0$) along the strip in the $m_{1/2}$--$m_0$
plane that is allowed by the dark matter constraints. These points
yield the following results for $\MW$ in the MSSM 
\begin{eqnarray}
&(1)& m_{1/2} = 250 \gev, m_0 = 1650 \gev, A_0 = -250 \gev \non \\
&& \Rightarrow \MW = 80.380 \gev,  \ \ \ \ \ \  
\MW^{\rm SM}(\MHSM = \Mh^{\rm MSSM})=80.361 \gev , 
\label{eq:focus1} \\[.5em]
&(2)& m_{1/2} = 330 \gev, m_0 = 2030 \gev, A_0 = -330 \gev \non \\
&& \Rightarrow \MW = 80.372\gev,\ \ \ \ \ \ 
\MW^{\rm SM}(\MHSM = \Mh^{\rm MSSM})=80.360 \gev ,
\label{eq:focus2} \\[.5em]
&(3)& m_{1/2} = 800 \gev, m_0 = 3685 \gev, A_0 = -800 \gev \non \\ 
&& \Rightarrow \MW = 80.361\gev,\ \ \ \ \ \ 
\MW^{\rm SM}(\MHSM = \Mh^{\rm MSSM})=80.359 \gev ,
\label{eq:focus3}
\end{eqnarray}
where the low-scale parameters of the MSSM have been obtained from the 
high-scale parameters $m_{1/2}$, $m_0$, $A_0$ with the help of the
program {\tt ISAJET~7.71}~\cite{isajet}. For comparison, the
corresponding prediction in the SM with $\MHSM = \Mh^{\rm MSSM}$ is
also given.

One can see from \refeqs{eq:focus1}--(\ref{eq:focus3}) that 
only for the point with the lowest $m_{1/2}$ value
a large difference of up to $\sim 20 \mev$ occurs in comparison to the
SM result. Such low $m_{1/2}$ values
correspond to very low masses of neutralinos and charginos. As a
consequence, the main contribution to the shift in $\MW$ arises from
the chargino and neutralino sector (see \refse{subsec:cndep}). Since 
for the $m_0$ value of \refeq{eq:focus1} the squarks are not completely
decoupled, there is also a small contribution from the squark sector.
Already with
slightly higher $m_{1/2}$ values, \refeq{eq:focus2}, the contribution
to $\MW$
becomes much smaller. For even higher values, \refeq{eq:focus3}, 
the resulting prediction for $\MW$ is very close to the
corresponding SM (decoupling) limit. These predictions deviate by about
$1.5 \si$ from the current 
experimental value of $\MW$, \refeq{MWexp}. The focus
point region therefore improves the prediction for $\MW$ only for very
low $m_{1/2}$. For most of the allowed parameter space, however, the
improvement is small. The deviation in $\MW$ contributes to the 
relatively bad fit quality of the focus point
region in a fit to electroweak precision observables as observed in 
\citere{ehow4}.


\subsubsection{Split SUSY}
\label{subsubsec:splitsusy}

Another scenario that has recently found attention is the so-called
``split-SUSY'' scenario~\cite{split}. Here scalar mass parameters are made
very heavy and only the fermionic masses (i.e.\ the chargino,
neutralino and gluino masses) are relatively small. According to the
analysis in the previous sections only a small deviation in the $\MW$
prediction from the SM limit is to be expected. 

\begin{figure}[htb!]
\begin{center}
\includegraphics[width=9cm,height=8.4cm]{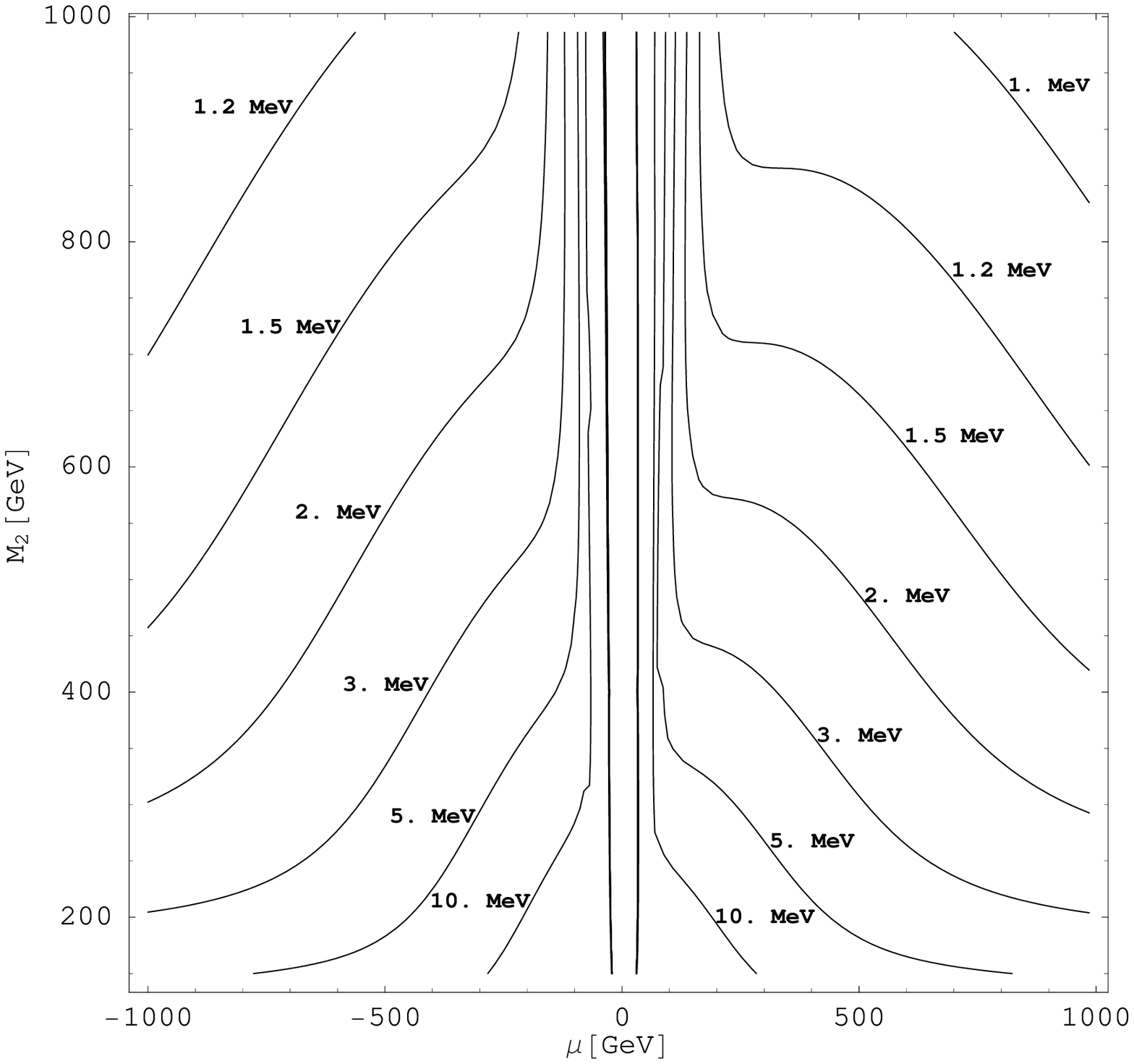}
\caption{Difference between the result for $\MW$ in the MSSM and the SM
  for large
  scalar fermion masses. The SUSY parameters are $\msusy = 3000\gev$,
  $A_{t,b}=2\,\msusy$, $\mgl=300 \gev$, $\MA=2000\gev$ and $\tb = 10$.}
\label{SplitSUSY}
\end{center}
\end{figure}

We have evaluated the prediction for $\MW$ in the split-SUSY
scenario. In \reffi{SplitSUSY} we show the SUSY contribution to $\MW$,
i.e.\ the deviation between the MSSM result in the split-SUSY scenario
and the corresponding SM result. This deviation is obtained by choosing
a large value for $\msusy$ and subtracting the SM result with 
$\MHSM = \Mh^{\rm MSSM}$. For definiteness we have chosen 
$\msusy = 3 \tev$ and $\MA = 2 \tev$. 
Choosing a higher scale for the scalar mass parameters would lead to
even slightly smaller deviations from the SM prediction than the ones
shown in \reffi{SplitSUSY}. The resulting shifts in $\MW$ are displayed in
\reffi{SplitSUSY} in the $\mu$--$M_2$ plane for $\tb = 10$.
The gluino mass has been set to $\mgl = 300 \gev$, however no visible
change in \reffi{SplitSUSY} occurs even for $\mgl = 3000 \gev$. 
As expected, only for rather small values of $\mu$ and $M_2$,
$M_2 \lsim 400 \gev$ and $|\mu| \lsim 500 \gev$ a deviation from the
SM limit larger than $5 \mev$ is found (a similar result has been
obtained in \citere{splitMW}, see also \citere{splitMW2}). Even with
the GigaZ precision 
for $\MW$ only a very light chargino/neutralino spectrum would result
in a $1 \si$ deviation in $\MW$ compared to the SM prediction. 

\subsubsection{MSSM parameter scans}
\label{subsubsec:scans}
Finally, we analyse the overall behaviour of $\MW$ in the MSSM by
scanning over a broad range of the SUSY parameter space. The following SUSY
parameters are varied independently of each other, within the given range, 
in a random parameter scan:\\

\begin{eqnarray}
 {\rm sleptons} &:& M_{{\tilde F},{\tilde F'}}= 100\dots2000\gev \non \\
 {\rm light~squarks} &:& M_{{\tilde F},{\tilde F'}_{\textup{up/down}}}
                   = 100\dots2000\gev \non \\
 \Stop/\Sbot {\rm ~doublet} &:& 
                         M_{{\tilde F},{\tilde F'}_{\textup{up/down}}}
                    = 100\dots2000\gev\non\\
 && A_{t,b} = -2000\dots2000\gev \non \\
 {\rm gauginos} &:& M_{1,2}=100\dots2000\gev \non \\
 && \mgl=195\dots1500\gev \non \\
 && \mu = -2000\dots2000\gev\non \\ 
 {\rm Higgs} &:& \MA=90-1000\gev \non \\
 && \tb = 1.1\dots60
\end{eqnarray}

We have taken into account the constraints on the MSSM parameter space
from the LEP Higgs searches~\cite{LEPHiggsMSSM,LEPHiggsSM} and the lower
bounds on the SUSY particle masses from \citere{pdg}.
Apart from these constraints no 
other restrictions on the MSSM parameter space were made.

\begin{figure}[htb!]
\vspace{1.0em}
\begin{center}
\includegraphics[width=10cm]{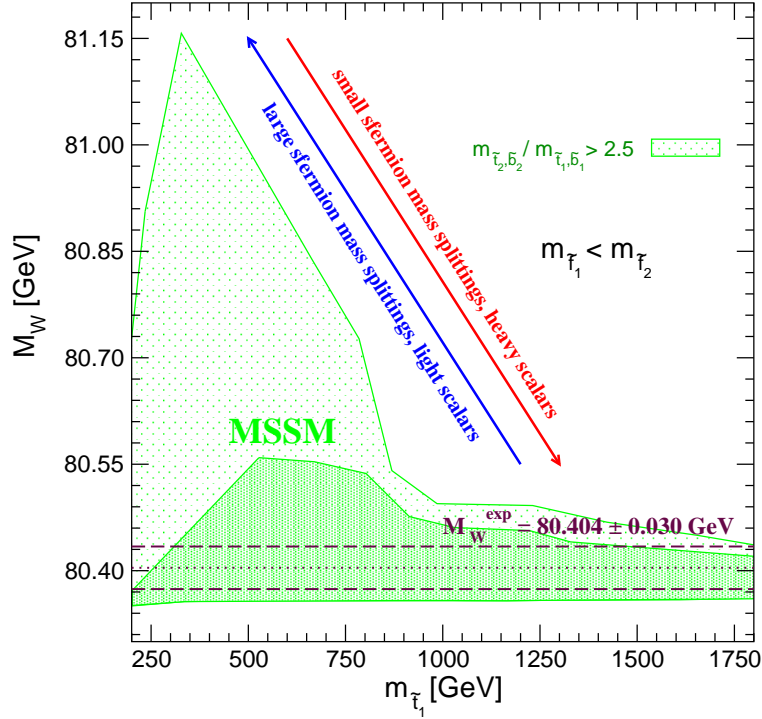}
\begin{picture}(0,0)
\CBox(-110,248)(-15,263){White}{White}
\end{picture}
\caption{Prediction for $\MW$ as function of $\mste$, the mass of the
  lighter stop squark. The SUSY parameters are varied independently of
  each other in a
  random parameter scan as described in the text.
  The top-quark mass is fixed at 
  its current experimental central value, 
  $m_t=172.5\gev$.} 
\label{MWMSf133scatter}
\end{center}
\end{figure}

In \reffi{MWMSf133scatter} we show the result for $\MW$ as a function
of the lightest $\Stop$~mass, $\mste$. The top-quark mass has been fixed 
to its current experimental central value, $\mt = 172.5 \gev$. 
The results are divided into a
dark (green) shaded and a light (green) shaded area. In the latter
at least one of the ratios $\mstz/\mste$ or $\msbz/\msbe$ exceeds~2.5,%
\footnote{
We work in the convention that $\msfe \le \msfz$.
}%
~i.e.\ the darker shaded region corresponds to a moderate splitting
among the $\Stop$~or $\Sbot$~doublets. In this region the MSSM
prediction for $\MW$ does not exceed values of about $80.550 \gev$.
In the case of very large
splitting in the $\Stop$~and $\Sbot$~doublets, much larger $\MW$ 
values up to $81.150 \gev$ would be possible (which are of course ruled 
out by the experimental measurement of $\MW$).

\begin{figure}[htb!]
\vspace{1.0em}
\begin{center}
\includegraphics[width=10cm]{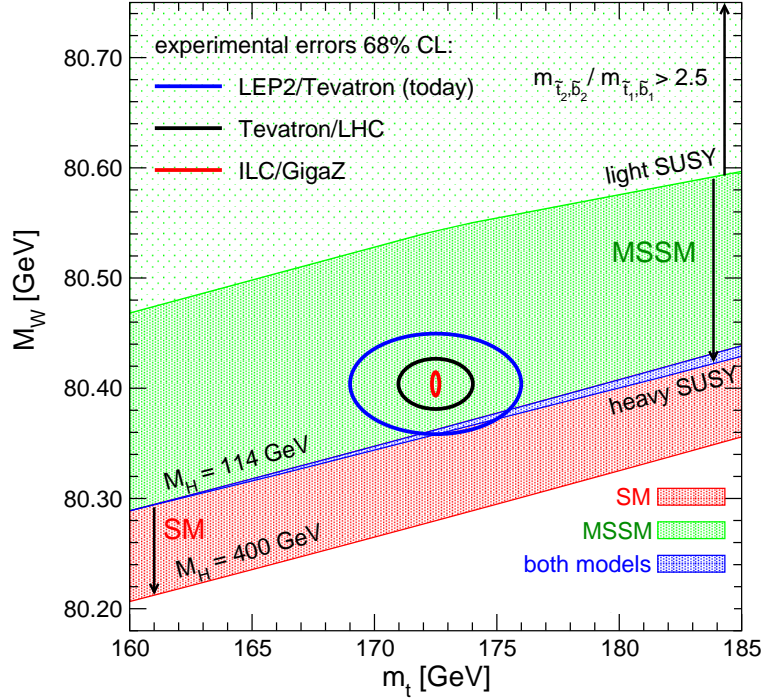}
\begin{picture}(0,0)
\CBox(-110,33)(-15,46){White}{White}
\end{picture}
\caption{Prediction for $\MW$ in the MSSM and the SM as a function of
$\mt$ in comparison with the present experimental results for $\MW$ and
$\mt$ and the prospective accuracies (using the current central values)
at the Tevatron / LHC and at the ILC. The allowed region in the MSSM,
corresponding to the light-shaded (green) and dark-shaded (blue) bands,
results from varying the SUSY parameters independently of each other
in a random parameter scan. The allowed region in the SM, corresponding
to the medium-shaded (red) and dark-shaded (blue) bands, results from
varying the mass of the SM Higgs boson from $M_H=114\gev$ to
$M_H=400\gev$. Values in the very light shaded region can only be
obtained in the MSSM
if at least one of the ratios $\mstz/\mste$ or $\msbz/\msbe$
exceeds~2.5.}
\label{fig:MWMT}
\end{center}
\end{figure}

In \reffi{fig:MWMT} we 
compare the SM and the MSSM predictions for $\MW$
as a function of $\mt$ as obtained from the scatter data. 
The predictions within the two models 
give rise to two bands in the $\mt$--$\MW$ plane with only a relatively small
overlap region (indicated by a dark-shaded (blue) area in \reffi{fig:MWMT}). 
The allowed parameter region in the SM (the medium-shaded (red)
and dark-shaded (blue) bands) arises from varying the only free parameter 
of the
model, the mass of the SM Higgs boson, from $\MHSM = 114\gev$, the LEP 
exclusion bound~\cite{LEPHiggsSM}
(upper edge
of the dark-shaded (blue) area), to $400 \gev$ (lower edge of the
medium-shaded (red) area).
The very light-shaded (green), the light shaded (green) and the
dark-shaded (blue) areas indicate 
allowed regions for the unconstrained MSSM. 
In the very light-shaded region (see \reffi{MWMSf133scatter}) 
at least one of the ratios $\mstz/\mste$ or $\msbz/\msbe$ exceeds~2.5,
while the decoupling limit with SUSY masses of \order{2 \tev}
yields the lower edge of the dark-shaded (blue) area. Thus, the overlap 
region between
the predictions of the two models corresponds in the SM to the region
where the Higgs boson is light, i.e.\ in the MSSM allowed region 
($\Mh \lsim 135 \gev$~\cite{feynhiggs,mhiggsAEC}). In the MSSM it
corresponds to the case where all 
superpartners are heavy, i.e.\ the decoupling region of the MSSM.
The current 68\%~C.L.\ experimental results%
\footnote{
The plot shown here is an update of 
\citeres{MWMSSM1LA,PomssmRep}.
}%
~for $\mt$ and $\MW$ are indicated in the plot. As can be seen from 
\reffi{fig:MWMT}, the current experimental 68\%~C.L.\ region for 
$\mt$ and $\MW$ exhibit a slight preference of the MSSM over the SM.
The prospective accuracies for the Tevatron/LHC 
($\de\mt^{\rm Tevatron/LHC} = 1 \gev$, 
$\de\MW^{\rm Tevatron/LHC} = 15 \mev$) and the ILC with GigaZ option 
($\de\mt^{\rm ILC/GigaZ} = 0.1 \gev$,  
$\de\MW^{\rm ILC/GigaZ} = 7 \mev$) 
are also shown in the plot (using the current
central values), indicating the 
potential for a significant improvement of the sensitivity of the
electroweak precision tests~\cite{gigaz}.

\subsection{Remaining higher-order uncertainties}
\label{sec:theounc}

As explained above, we have incorporated all known SM corrections into
the prediction for $\MW$ in the MSSM, see \refeq{eq:obsSMSUSY}. This
implies that the theoretical uncertainties from unknown higher-order
corrections reduce to those in the SM in the decoupling limit. In the
SM, based on all higher-order contributions that are currently known,
the remaining uncertainty in $\MW$
has been estimated to be~\cite{MWSM}
\begin{equation}
\de\MW^{\rm SM} = 4 \mev~.
\label{eq:SMunc}
\end{equation}
Below the decoupling limit an additional
theoretical uncertainty arises from higher-order corrections involving
supersymmetric particles in the loops. 
This uncertainty 
has been estimated in \citere{drMSSMal2B} for the MSSM with real
parameters depending on the overall
sfermion mass scale $\msusy$, 
\begin{eqnarray}
\de\MW &=& 8.5 \mev \mbox{ for } \msusy < 500 \gev , \non \\
\de\MW &=& 2.7 \mev \mbox{ for } \msusy = 500 \gev ,  \label{eq:MWunc}\\
\de\MW &=& 2.4 \mev \mbox{ for } \msusy = 1000 \gev . \non 
\end{eqnarray}
The full theoretical uncertainty from unknown higher-order corrections
in the MSSM with real parameters can be obtained by adding in 
quadrature the SM uncertainties from \refeq{eq:SMunc} and the
SUSY uncertainties from \refeq{eq:MWunc}. This yields 
$\de\MW=(4.7\,-\,9.4)\mev$
depending on the SUSY mass scale~\cite{drMSSMal2B}.

Allowing SUSY parameters to be complex adds an additional theoretical
uncertainty from unknown higher-order corrections to the $\MW$ prediction. 
While at the one-loop level
the full complex phase dependence is included in our evaluation, it is
only approximately taken into account at the two-loop level 
as an interpolation between the known results for the phases 0 and
$\pi$, see \refeq{phases2L}.
We concentrate here on the complex phase in the scalar top sector,
$\phi_{A_t}$ (we keep $\phi_\mu$ fixed), since our analysis above has 
revealed that the impact of
the other phases is very small already at the one-loop level.

We estimate the uncertainty from unknown higher-order corrections
associated with the phase dependence as follows.
The full result for $\MW(\phi)$, i.e.\ for a given phase $\phi$ (where
the two-loop corrections are taken into account as described in
\refse{subsubsec:twoloop}) lies by construction in the interval 
\begin{equation}
[ \MW^{\rm full}(0)   + (\MW^{1 {\rm L}}(\phi) - \MW^{1 {\rm L}}(0)), \quad
   \MW^{\rm full}(\pi) + (\MW^{1 {\rm L}}(\phi) - \MW^{1 {\rm L}}(\pi)) ]~.
\end{equation}
The minimum difference of $\MW^{\rm full}(\phi)$ to the boundary of
this interval,
\begin{equation}
{\rm Min}\{ 
|\MW^{\rm full}(\phi) - (   \MW^{\rm full}(0) 
              + (\MW^{1 {\rm L}}(\phi) - \MW^{1 {\rm L}}(0)))|,
|\MW^{\rm full}(\phi) - (   \MW^{\rm full}(\pi) 
              + (\MW^{1 {\rm L}}(\phi) - \MW^{1 {\rm L}}(\pi)))| \}~,
\end{equation}
can be taken as estimate for the theoretical uncertainty
(this automatically ensures that no additional uncertainties arise for 
$\phi = 0, \pi$). 

As representative SUSY scenarios we have chosen SPS~1a,
SPS~1b, and SPS~5, each for $\msusy = 1000\gev$, $500\gev$, and for
$\msusy<500\gev$~%
\footnote{
The lowest values considered for $\msusy$ are roughly 300, 300, 400 $\gev$
for SPS~1a, SPS~1b, SPS~5, respectively. 
For lower values the parameter points are 
excluded by Higgs mass constraints. The light stop mass for the SPS~5
point lies considerably below $400\gev$.
}%
~(as above we have varied $\msusy$, $A_{t,b}$ and $\mu$ using a common
scale factor). A non-zero complex phase $\phi_{\At}$ has been introduced
as an additional parameter. In order to arrive at a conservative
estimate for the intrinsic error, we use the value obtained for 
$\phi_{\At} = \pm \pi/2$.
In this way we obtain
\begin{eqnarray}
\de\MW &=& 3.2 \mev \mbox{ for } \msusy < 500 \gev , \non \\
\de\MW &=& 2.0 \mev \mbox{ for } \msusy = 500 \gev ,  \label{eq:MWunc2}\\
\de\MW &=& 0.7 \mev \mbox{ for } \msusy = 1000 \gev . \non 
\end{eqnarray}
The full theoretical uncertainty from unknown higher-order corrections
in the MSSM with complex parameters can
now be obtained by adding in 
quadrature the SM uncertainties from \refeq{eq:SMunc}, the theory
uncertainties from \refeq{eq:MWunc} and the additional 
SUSY uncertainties from \refeq{eq:MWunc2}. This yields 
$\de\MW=(4.7 - 9.9)\mev$ depending on the SUSY mass scale.

The other source of theoretical uncertainties besides the one from
unknown higher-order corrections is the parametric uncertainty induced
by the experimental errors of the input parameters. The current
experimental error of the top-quark mass~\cite{newestmt} induces the
following parametric uncertainty in $\MW$
\begin{equation}
\de\mt^{\rm exp} = 2.3 \gev \;\; \Rightarrow \;\; 
\de\MW^{{\rm para}, \mt} = 14 \mev~,
\end{equation}
while the uncertainty in 
$\De\al_{\rm had}^{(5)}$~\cite{delDelalphahad} results in 
\begin{equation}
\de(\De\al_{\rm had}^{(5)}) = 3.6 \times 10^{-4} \;\; \Rightarrow \;\; 
\de\MW^{{\rm para}, \De\al_{\rm had}^{(5)}} = 6.5 \mev~.
\end{equation}
The uncertainty in $\mt$ will decrease during the next years as a
consequence of  
a further improvement of the accuracies at the Tevatron and the
LHC. Ultimately it will be reduced by more than an order of magnitude at
the ILC~\cite{deltamt}. For $\De\al_{\rm had}^{(5)}$ one can hope for
an improvement down to $5 \times 10^{-5}$~\cite{fredl}, reducing the
parametric uncertainty to the $1 \mev$ level (for a discussion of the
parametric uncertainties  
induced by the other SM input parameters see e.g.\ \citere{PomssmRep}).
The effect of $\de(\De\al_{\rm had}^{(5)})$ on \reffi{fig:MWMT} is small.
In order to reduce the theoretical uncertainties from unknown
higher-order corrections to the $1 \mev$ level, further results on SM-type
corrections beyond two-loop order and higher-order corrections involving
supersymmetric particles will be necessary.


\section{Conclusions}
\label{sec:conclusions}

We have presented the currently most accurate evaluation of the
$W$~boson mass in the MSSM. The calculation includes the complete
one-loop result, taking into account for the first time the full
complex phase dependence,  and
all known higher-order corrections in the MSSM. Since the evaluation of
higher-order contributions in the SM is more advanced than in the MSSM, 
we have incorporated all available SM corrections which go beyond the 
results obtained so far in the MSSM. Our prediction for $\MW$ in the
MSSM therefore reproduces the currently most up-to-date SM
prediction for $\MW$ in the decoupling limit where the masses of all
supersymmetric particles are large. A public computer code based on our
result for $\MW$ is in preparation.

We have analysed in detail the impact of the various sectors of the MSSM 
on the prediction for $\MW$, focussing in particular on the dependence on the
complex phases entering at the one-loop level. The most pronounced phase
dependence occurs in the stop sector, where the effect of varying the
complex phase that enters the off-diagonal element in the stop mass
matrix can amount to a shift of more than 20~MeV in $\MW$. It should be
noted, however, that the complex phases in the squark sector at
the one-loop level enter only via modifications of the squark masses and
mixing angles. As a consequence, a precision measurement of the
$\cp$-conserving observable $\MW$ alone will not be sufficient to reveal
the presence of $\cp$-violating complex phases. The phase dependence of
$\MW$ will however be very valuable for constraining the SUSY parameter
space in global fits where all accessible experimental information is
taken into account.

We have illustrated the sensitivity of the precision observable $\MW$ to
indirect effects of new physics by comparing our MSSM prediction with
the SM case. Confronting the MSSM prediction in different SUSY scenarios
and the SM result as a function of the Higgs-boson mass with the current
experimental values of $\MW$ and $\mt$, we find a slight preference for
non-zero SUSY contributions. As representative SUSY scenarios we have
studied various SPS benchmark points, where we have varied all mass
parameters using a common scalefactor. The MSSM prediction lies within 
the $1\si$ region of the experimental $\MW$ value if the SUSY mass scale is
relatively light, i.e.\ lower than about 600~GeV. 

The prospective improvement in the experimental accuracy of $\MW$ at the
next generation of colliders will further enhance the sensitivity to
loop contributions of new physics.
We have found that even for a SUSY mass scale of
several hundred GeV the loop contribution of supersymmetric particles to
$\MW$ is still about 10~MeV, which may be probed in the high-precision
environment of the ILC. In SUSY scenarios with even higher mass scales,
however, it is unlikely that SUSY loop contributions to $\MW$
can be resolved with the currently foreseen future experimental accuracies. 
We have studied in this context the focus point and the split-SUSY
scenarios, which have recently received significant 
attention in the literature. In the focus point scenario only at
the lower edge of the allowed parameter region in the $m_{1/2}$--$m_0$
plane a sizeable
contribution to $\MW$ can be achieved. For higher values, as well as for
the split-SUSY scenario, we find that the SUSY
loop effects are in general very small, so that it is not possible to
bring the prediction for $\MW$ significantly closer to the current
experimental central value as compared to the SM case.

Finally we have analysed the theoretical uncertainties in the $\MW$
prediction that arise from the incomplete inclusion of the complex
phases at the two-loop level. We estimate that this uncertainty can 
amount up to roughly $3 \mev$ in the prediction for $\MW$,
depending on the SUSY mass scale. Combined with the estimate of the
possible effects of other unknown higher-order corrections we 
find that the total uncertainty from unknown higher-order
contributions is currently about $10 \mev$ for small SUSY mass scales.


\subsection*{Acknowledgements}

We are grateful to T.~Hahn for many helpful discussions on various
aspects of this work. S.H.\ and G.W.\
thank the Max Planck Institut f\"ur Physik, M\"unchen, for kind hospitality 
during part of this work.




\end{document}